\documentclass{article}

\usepackage{arxiv}

\usepackage[utf8]{inputenc}
\usepackage[T1]{fontenc}
\usepackage{listings}

\usepackage{amsmath, amsfonts, amssymb}
\usepackage{graphicx}
\graphicspath{{figures/}}
\usepackage{booktabs}
\usepackage{array}
\usepackage{longtable}
\usepackage{microtype}
\usepackage{xcolor}
\usepackage[colorlinks=true, linkcolor=blue, citecolor=blue, urlcolor=blue]{hyperref}
\usepackage[round, authoryear]{natbib}

\title{(Whose defaults?) Is artificial intelligence reorienting archaeological methods?}

\renewcommand{\shorttitle}{Is artificial intelligence reorienting archaeological methods?}

\author{
  \textbf{Lorenzo Cardarelli} \\
  Seminar f\"ur Ur- und Fr\"uhgeschichte \\
  Georg-August-Universit\"at G\"ottingen \\
  G\"ottingen, Germany \\
  \href{https://orcid.org/0000-0002-2436-9967}{\texttt{ORCID: 0000-0002-2436-9967}} \\
  \And
  \textbf{Roberto Ragno}\thanks{Corresponding author: \href{mailto:rr673@cam.ac.uk}{\texttt{rr673@cam.ac.uk}}} \\
  McDonald Institute for Archaeological Research \\
  University of Cambridge \\
  Cambridge, UK \\
  \href{https://orcid.org/0000-0002-7333-9035}{\texttt{ORCID: 0000-0002-7333-9035}} \\
}

\date{}

\begin{document}
\addtocontents{toc}{\protect\setcounter{tocdepth}{-1}}
\maketitle

\begin{abstract}
The paper asks whether generative AI and `vibe coding' are narrowing the range of computational methods used in archaeological research, and tests this along two lines. We analysed roughly 119,000 archaeology abstracts indexed in Scopus for 2010--2025 and used a locally run large language model to extract and cluster the techniques each reports (25 operative categories, 241 data-driven clusters). A Bayesian hierarchical Dirichlet-multinomial model finds a small but credible compositional shift after 2023, roughly 2.35 times smaller than the field's existing heterogeneity; no single technique changed credibly, and the effective number of methods rose from 87.6 to 111.2. Separately, two open-weight models asked to recommend methods for 28 standardised research problems, under novice, intermediate, and expert guidance, produced recommendations about a third as diverse as the literature, and a quarter under novice prompting; a negative binomial regression shows that a method's pre-2023 prevalence predicts how often it is recommended, most strongly for the novice profile. The results are consistent with a convergent pressure on method choice that remains latent rather than visible in the published record, though the design cannot establish causation.
\end{abstract}


\section{Introduction}\label{sec:intro}

The public release of ChatGPT at the end of 2022 marked a turning point in the relationship between researchers and computational systems: for the first time, natural language became the primary interface for instructing a machine. Since then, a rapid succession of increasingly capable models has reshaped how research is conducted, from data collection to the formalisation and execution of analyses \citep{Zhang2025exploring, Hao2026artificial, Traberg2026ai, Underwood2025impact}. The `trajectory of use' described by Zhang et al. (2025) has, in practice, become an increasingly progressive delegation process, developing from simple conversational assistance and manual code execution, to fully autonomous \emph{agentic} workflows. These models can autonomously plan, write and execute \citep{Sapkota2025vibe}. The use of Large Language Models (LLMs) is already visible in the archaeological literature \citep{Qi2025large}. Indeed, our preliminary analysis of approximately 120,000 Scopus abstracts spanning 2010 to 2025 shows a marked increase in linguistic markers associated with LLM-assisted writing from 2022 onwards, in line with what \citet{Liang2025monitoring} already observed for STEM preprints.

\begin{figure}[htbp]
\centering
\includegraphics[width=0.85\linewidth]{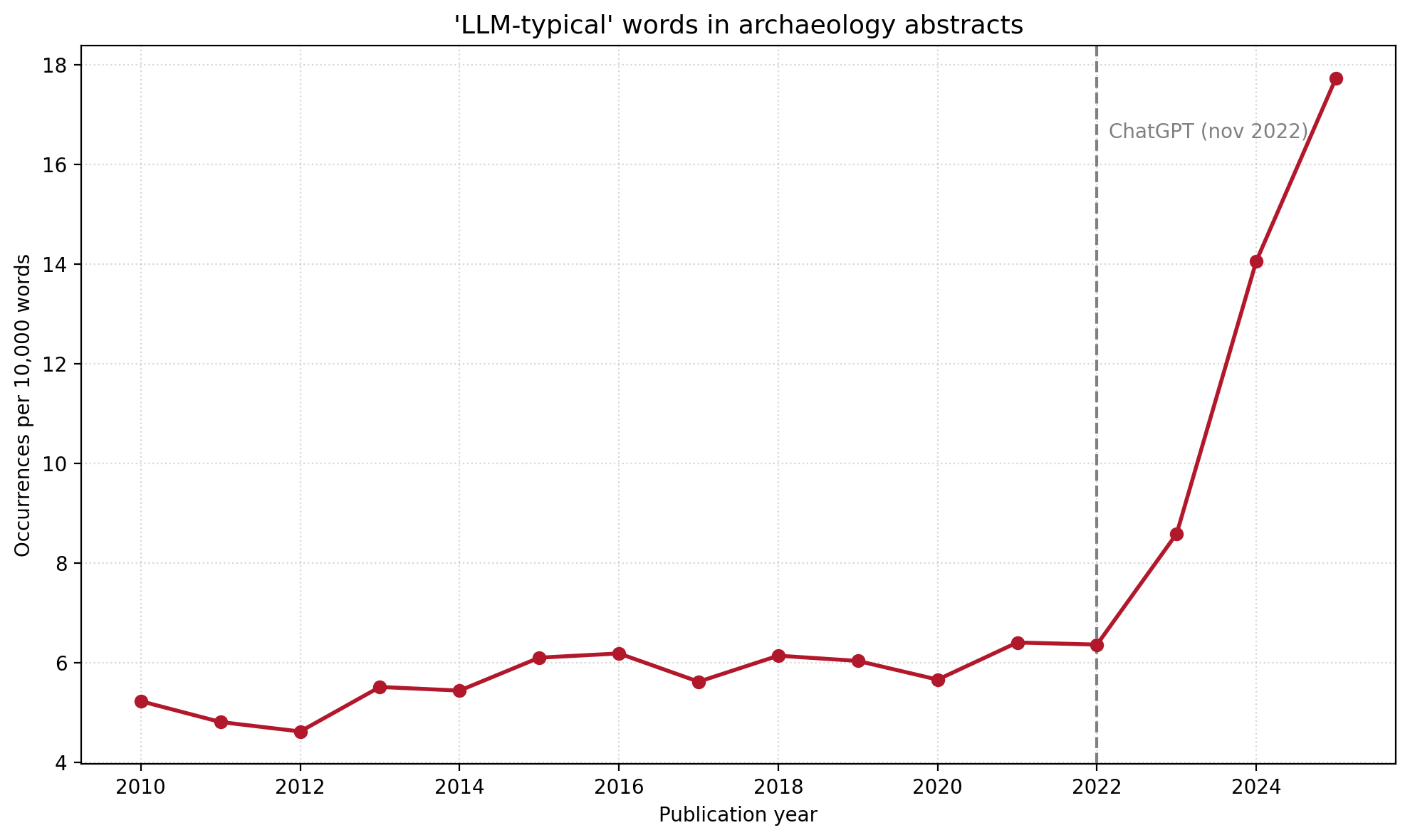}
\caption{Frequency of `LLM-typical' words in archaeological literature (2010--2025). Occurrences per 10,000 words across approximately 120,000 Scopus abstracts. The tracked vocabulary comprises 25 marker words identified as characteristic of LLM-generated text (\emph{delve, intricate, intricacies, showcase, underscore, pivotal, realm, tapestry, meticulous, boast, leverage, nuanced, multifaceted, encompass, comprehensive, notably, crucial, garner, foster, seamless, robust, paradigm, holistic, elucidate, endeavor}). The vertical dashed line marks the release of ChatGPT (November 2022).}
\label{fig:llm_words}
\end{figure}

Logic and programming represent the privileged domains for the development of this type of Artificial Intelligence (AI), which is predominantly text-based. It is no coincidence that many of the benchmarks used to measure a model's efficiency in the industry's race towards Artificial General Intelligence (AGI) (such as SWE-bench or LiveCodeBench \citep{Jimenez2024swebench} consist of complex programming tests to evaluate its autonomous planning and efficiency. These are syntactic and logical metrics of a binary nature, where code admits no ambiguity: it either compiles or it fails. 

The integration of these tools in archaeology is not only a practical issue. Although the discipline routinely embraces computational approaches (e.g. databases, GIS, 3D modelling, machine learning), archaeology remains an intrinsically interpretative field whose object of study is, by definition, complex, fragmentary, and partial \citep{Gattiglia2025managing}. The inductive and empirical reasoning required for reconstructing the past bears little resemblance to the binary logic of computational benchmarks. There is a paradox here: the computational archaeologist and the software engineer now share the very same agentic interface for generating code \citep{OBrien2026survey, VanNoorden2023ai}, yet their epistemic goals remain radically different. Code is not a neutral tool but the \emph{epistemic infrastructure} through which archaeological data are filtered, analysed, classified, visualised, and interpreted \citep{Gattiglia2022postphenomenological, Gattiglia2025managing}. Every algorithm encodes assumptions about what constitutes meaningful variation, and every pipeline embodies decisions about which information to preserve and which to discard. In Don Ihde's terms, code functions as a `hermeneutic relation' with the world \citep{Ihde1990technology}: it does not simply transmit archaeological reality to the researcher but actively mediates and transforms it \citep{Huggett2021algorithmic}.

When the authorship of this mediating layer is delegated to an AI, specifically a LLM,  the epistemological implications run deep. This concern is not new to the discipline. At every major methodological turn, computational archaeology has grappled with analogous tensions: the quantitative revolution of the 1960s, with \citet{Clarke1968analytical}'s analytical archaeology and \citet{Binford1962archaeology, Binford1968perspectives}'s nomothetic programme, brought computational tools to the centre of the discipline, but in doing so they embedded theoretical assumptions about what constitutes meaningful variation in human behaviour. It was precisely these assumptions that the post-processualist critique set out to deconstruct. \citet{Hodder1991reading} and \citet{Shanks1987reconstructing} insisted that every classification, model, and statistic is situated, culturally conditioned, and theory-laden. This idea was triggered, to a large extent, by the opacity of the computational methods that processualism had normalised. Subsequent technological waves raised similar questions without displacing the researcher as the epistemic subject. GIS in the 1990s raised the problem of the proprietary black-box \citep{Wheatley2002spatial}, Bayesian chronological modelling imposed a formal language many used without mastering its probabilistic foundations \citep{Bayliss2007bayesian}, and machine learning for ceramic classification and remote sensing \citep{Bellat2025machine, Bickler2021machine} pushed the boundary further still. \citet{Kristiansen2024history}'s longer historical account of interdisciplinarity in archaeology describes it as a recurring pendulum swinging between science-based and humanistic-based interpretative dominance, with oscillations at intervals of roughly 30--50 years: each science-driven `revolution' triggers, after a phase of implementation, a `counterrevolution' from cultural-historical and interpretive positions. The quantitative revolution and its post-processualist backlash outlined above are, in this reading, simply the discipline's second documented turn of this cycle. \citet{Kristiansen2014towards} described this juncture as a `Third Science Revolution': the convergence of big data, genomics, high-resolution dating and advanced computation engaged in conversation with humanistic interpretation, rather than positioned in opposition to it. However, in practice, successive data-driven developments in archaeology have often relegated reflexivity to a secondary role. This has produced what \citet{Niklasson2014shutting} aptly characterised as an `add critics and stir' approach, in which critical engagement is appended after the fact rather than built into the analytical framework from the outset.   
This pattern is structural rather than accidental: quantitative revolutions in archaeology repeatedly promised dialogue between data and interpretation, but made data the priority and interpretation the by-product \citep{Chilton2014plus, GonzalezRuibal2014archaeological}. All these precedents have one thing in common and distinguish themselves from what lies ahead: every tool, however complex, requires the user to retain at least a functional understanding of the underlying logic. For example, anyone using a GIS had to understand what a buffer or a viewshed was, and anyone fitting a Bayesian model had to grasp the concept of a prior. Similarly, anyone training a neural network had to define an architecture, hyperparameters and a validation set. Agentic AI and \emph{Vibe coding} (\S\ref{sec:background}) break this continuity: for the first time, a researcher can produce an entire, functioning analytical tool without understanding any of the computational operations that compose it. The distance between intent and implementation, which every previous innovation had progressively narrowed, is now closed at the operational level (\emph{technē}), yet simultaneously rendered insurmountable at the epistemic level  (\emph{epistēmē}). Vibe coding is not creating the conditions for the asymmetry between data and reflexivity identified by Kristiansen's critics, but it radicalises it to a degree that no previous tool has come close to achieving.

At this point it is worth laying our cards on the table. As early-career computational archaeologists early, we write advocating for the integration of these tools in our programming workflows, under firm conditions (which will be discussed in Section~\ref{sec:discussion}). Vibe coding has enabled us to build in hours what would previously have taken weeks: custom pipelines, responsive interfaces and analytical tools that would have been impossible to develop without a dedicated software team. This is a significant development for researchers accustomed to working independently with limited time and resources. Our critique does not stem from a distrust of technology, but rather from the opposite. It is precisely because we have experienced this transformative potential first-hand that we feel compelled to examine it honestly. This examination begins with an honest acknowledgement that the democratising potential of these tools is real but not unconditional. Access issues, cost, data privacy and the uneven distribution of computational infrastructure complicate any simple narrative of empowerment, particularly for early-career researchers and didactic activity. We explore these issues in the next section, before conducting an empirical investigation. 

\subsection{Background and Theoretical Framework}\label{sec:background}

The history of programming is one of increasing abstraction. Starting with assembly and moving on to C, Python, and the myriad programming languages that now exist, each step has narrowed the cognitive distance between human ideas and their execution. \emph{Vibe coding}, a term coined by Andrej Karpathy (cofounder of OpenAI) in early 2025, is the latest leap: a workflow in which the operator expresses intent in natural language, leaving the LLM to manage the implementation of syntax, logic and architecture entirely. The technical debate distinguishes between several variants. \emph{Vibe engineering,} as defined in \citet{Bamil2025vibe}, is a more disciplined approach that combines LLMs with code review, testing and documentation. \emph{Agentic coding} involves systems in which the AI autonomously plans and executes entire workflows, relegating the human to the role of supervisor \citep{Bamil2025vibe, Sapkota2025vibe}. While these are real distinctions marking different degrees of delegated autonomy, the conceptual core remains identical: in every case, the researcher forfeits responsibility for the formal and logical construction of the analytical tool.

For archaeologists (and researchers more broadly) who work without institutional software teams or dedicated budgets and often independently, this is a significant development. Vibe coding has lowered the barriers for creating custom analytical tools, including pipelines, interfaces, and visualisations, which were previously accessible only to those with formal programming training or specialist collaborative partners. This democratising potential is real, and any honest engagement with the technology must acknowledge it. The difficulty stems from the fact that this potential is accompanied by a set of challenges that are amplified, rather than resolved, in a research context. \citet{Bamil2025vibe} identifies six of these challenges: (1) intent alignment and ambiguity (since translating vague descriptions into code is hard by its very nature); (2) reproducibility and consistency (since stochastic generative models can produce different results for the same prompt); (3) bias and ethics (since training data carry social biases that can be amplified by informal descriptors); (4) explainability and transparency (since the generated code is difficult to understand); (5) maintainability (since the code may not follow established patterns); and (6) security (since AI-generated code often has vulnerabilities). In a research setting, logic errors are more difficult to detect than errors in generated prose. If a script runs and returns output in the expected format, subtle mistakes that distort results in non-obvious ways can go unnoticed. 

A reasonable counterargument is that reproducible scientific code already relies on libraries that are opaque to most users. For example, running a kernel density estimate in QGIS, is also a kind of black-box process. As we argue below, the difference is one of kind rather than degree. The theoretical core of the problem lies in the nature of code as an epistemic artefact. Writing code is not a mechanical translation process, but rather a modelling process in which the archaeologist's implicit knowledge (the \emph{know-how} bound up with materials, instruments, and historical questions) is progressively articulated into explicit logical structures. In Polanyian terms \citep{Polanyi1966logic}, traditional programming forces researchers to formalise their tacit knowledge. This entails making continuous micro-decisions about category boundaries, how to treat missing data and statistical weights. The promise of vibe coding is to relieve the researcher of this cognitive burden. 

Here, however, we draw the distinction that constitutes the central theoretical contribution of this paper: the difference between \emph{externalising} and \emph{bypassing} knowledge. When archaeologists use an established R package or a GIS, they externalise the operational calculation while retaining logical control over the procedure. In vibe coding, by contrast, the researcher bypasses the formalisation phase altogether, and the tool ceases to be a medium \emph{through which} the archaeologist thinks, becoming instead a substitute that thinks \emph{in their place}. The difference is concrete. Let us consider two different prompts. `What types can you identify in this dataset?' delegates the entire categorical process, including the choice of meaningful variables, the distance metric and the threshold between types, leaving the logic of classification opaque. `Create a K-means script using diameter as a feature, with k = 3, and return the centroids' formalises the interpretative assumptions and uses the model as an implementation accelerator. In the first case, the code bypasses cognition; in the second, it externalises it.

\citet[p.~162]{Wylie2002thinking} described archaeological arguments as \emph{cables} woven from many independent strands of evidence. LLM-generated code threads these opaque strands together to form these cables. When researchers do not understand the clustering logic of an autonomously generated spatial analysis, the strength of the cable no longer rests on methodological robustness, but on \emph{trust} in the statistical correctness of the model. Consequently, scientific accountability is undermined. Two systemic risks follow from this, and both are directly relevant to the empirical investigation that follows. The first is a \emph{secondary black-box}: adds a second layer to the model's native opacity \citep{Fawzy2025vibe, Sarkar2025vibe}. This results in researchers producing working software whose internal logic is incomprehensible to them, thereby nullifying the principles of open science. The second risk is \emph{algorithmic agency} \citep{Huggett2021algorithmic}. Generative systems are trained on the aggregate of existing scientific production and, by design, tend towards statistical `averaging' of that distribution: a hermeneutic \emph{mean collapse} in which the most common methods are favoured over potentially more appropriate, less common alternatives.  For archaeology, the concrete consequence would be the progressive homogenisation of computational methods reported in the literature, as researchers would follow the model's defaults rather than the logic of their research problem. The present study aims to empirically test the \emph{mean-collapse hypothesis}.

Algorithmic agency is not only a structural risk to the published record; it is also an educational risk. This is where the democratisation narrative requires its most serious revision. The researchers currently entering the discipline are the first generation for whom LLM assistance has been available from the outset of their training. In ten to fifteen years, the question will not be whether they can write code, but whether they can evaluate what the code does, recognise when an output is methodologically inappropriate and push back against a model's defaults with a theoretically grounded alternative. There is accumulating evidence that delegating cognitive tasks before consolidating the underlying competence has measurable effects on learning and critical thinking \citep{Gerlich2025ai, Si2026thinking}. The mechanism that determines which questions can be asked today is the same one that reduces the ability to ask different questions tomorrow. 

This educational risk is exacerbated by the uneven distribution of access outlined above. While a standard consumer subscription to a frontier model costs roughly the same as a streaming service and is affordable for most researchers in medium- to high-income contexts, research workflows quickly outgrow consumer tiers. API access costs many times more \citep{StokelWalker2026ai}, and locally hosted alternatives (often the only option when working with sensitive data such as genetic sequences or unpublished archival material) require high-end GPU infrastructure that many institutions, particularly in the Global South, cannot provide. This trajectory is already evident in projects such as ARIS \citep{Ghareeb2026aris, Yang2026aris}, a multi-agent system capable of conducting scientific research autonomously, from generating hypotheses to designing experiments to drafting manuscripts. Such systems operate exclusively through API access at scale, which places them firmly beyond the reach of researchers without substantial institutional backing \citep{StokelWalker2026ai}. This raises the prospect of a self-reinforcing cycle in which those with the resources to deploy autonomous research infrastructure publish more, attract more funding and thereby acquire the means to publish even more. This can be seen as a `Matthew effect' \citep{Crowson2025academic}: those who are well-equipped can take full advantage of these tools, while those who are not are limited to the free tiers. Those most dependent on LLM defaults because they lack the prior methodological knowledge to constrain the output are precisely those with the least access to the tools that would make those defaults most navigable. 

None of this is an argument against the use of these tools. Rather, it is an argument for using them with a degree of intentionality that the current pace of adoption makes difficult to sustain, and for the kind of critical engagement that the rest of this paper attempts to demonstrate.

\section{Aims}\label{sec:aims}

This study pursues two interconnected aims. The first is empirical: we test the mean-collapse hypothesis, according to which uncritical use of LLMs narrows the range of computational methods used in archaeological research. We evaluate this through two complementary lines of evidence: (i) the distribution of computational methods across approximately 120,000 archaeological abstracts indexed in Scopus, to establish whether this range shifted after 2023 (since ChatGPT was released in November 2022, we used 2023 as a cut-off point for its use) beyond existing disciplinary trends; and (ii) a direct query of local LLMs, asked to recommend methods for a set of standardised archaeological research problems under varying degrees of methodological guidance, comparing the concentration of their recommendations against the diversity observed in the published literature. From this investigation we expect different possible outcomes.  If homogenisation is already evident in both the LLM outputs and the literature, this would suggest that its effect is already shaping archaeological research. However, if homogenisation is confined to the models alone, this would point to a structural bias that has not yet propagated into published practice. In this case, we argue that stronger AI governance from academic publishers is needed to prevent this from happening.

The second aim is reflective. Building on these quantitative findings, we open a broader discussion of the epistemological implications of vibe coding in archaeology, proposing a framework that neither rejects nor uncritically accepts these tools.

\section{Materials and Methods}\label{sec:methods}

\subsection{Dataset and Bibliometric Stream}\label{sec:dataset}

The dataset used in this study was retrieved from Scopus (Elsevier). Starting from the Scopus Source List (\url{https://www.scopus.com/sources}), all journals including the code 1204 All Science Journal Classification Codes - ASJC) (Archaeology) were identified.  This code enables the identification of the Scopus Source IDs, which were then used in a Python script to download the entire dataset. Data collection was performed programmatically via a custom Python script accessing the Scopus Search API. Records were retrieved for the publication years 2010 to 2025, capturing the period during which computational approaches gained traction in the discipline (2010--2022), which also provides a statistical baseline for identifying authentic trends, as well as the period during which LLMs coexisted with these established approaches (2023--2025). Due to the proprietary nature of Scopus and the licensing restrictions imposed by Elsevier, the raw metadata dataset cannot be made publicly available. The Scopus Source IDs that define the corpus are likewise Elsevier identifiers and are not redistributed. However, to ensure full methodological transparency and reproducibility, the custom Python script, the exact query parameters, and the deterministic procedure for regenerating the identical list of 546 source identifiers from Elsevier's publicly downloadable Scopus Source List are provided in \S{}S3 of the ESM.

For each journal article, a wide range of metadata fields was extracted, including title, authors, and year. For the complete list of the 26 extracted metadata fields, refer to Table S3.2 in the ESM.

Prior to analysis, the dataset was deduplicated using the Elsevier Identifier (EID) as the primary key. The final cleaned dataset, from which records lacking an abstract or authors were removed, was exported in both CSV and XLSX formats for downstream analysis. The resulting dataset includes 119,327 journal articles.

\subsection{Method Classification and Taxonomy Creation}\label{sec:taxonomy}

In order to identify the computational and statistical techniques reported across the corpus, each abstract was processed through an automated information extraction pipeline based on a LLM running entirely on local hardware\footnote{Analyses were conducted on a computer equipped with 13th Gen Intel i9-13900K, RTX A4000. The full software environment, and the model, quantisation, context window, temperature and seed used at each stage of the pipeline, are given in \S{}S1 and \S{}S2 of the ESM.}. The extraction was performed using Qwen3.5-9B, a 9-billion-parameter instruction-tuned language model \citep{Yang2025qwen3}. The model was executed with full GPU offloading and a context window of 4,096 tokens, which accommodates any single abstract together with the system prompt. The model was queried through a structured chat-completion interface following a strict zero-shot prompt design. The system prompt instructed the model to act as a scientific classifier and identify all the computational and statistical techniques mentioned in a given abstract. The following constraints were observed: (i) techniques were labelled using their standard, widely recognised naming convention (e.g. \emph{Random Forest}, \emph{PCA}, \emph{LSTM}, \emph{Kriging}); (ii) the level of specificity had to match what was explicitly stated in the abstract. Broad labels such as `deep learning' were only permitted when no more specific method was mentioned. The model was explicitly prohibited from inferring unstated techniques; (iii) software packages and programming languages (e.g. Python, R, QGIS) were excluded; (iv) when multiple methods were present, they were to be listed separated by a pipe character (|); and (v) if no computational method was identifiable, the model was instructed to return the string \emph{None}. Few-shot output examples were embedded in the system prompt to anchor the expected response format and minimise free-form generation. The prompt is reproduced in \S{}S4.1 of the ESM. To maximise output determinism, temperature was set to 0.01 \citep{Li2025exploring} and a fixed random seed (42) was applied both at model initialisation and at each generation call. This extraction strategy was preferred over document-level topic modelling approaches such as BERTopic or LDA \citep{Blei2003latent, Grootendorst2022bertopic}, which assign a single dominant topic (or a unimodal topic distribution) to each document. This assumption would be structurally incompatible with our task: a single abstract routinely reports multiple methodologically heterogeneous techniques (e.g. a spatial interpolation method combined with a supervised classifier and a dimensionality reduction step), each of which constitutes an independent unit of analysis. The LLM extraction approach, by contrast, is designed to enumerate all explicitly stated techniques regardless of their number or mutual distance in the embedding space. This approach allows us to identify 8,404 papers with at least one computational method.

The raw method strings extracted by the LLM in the previous step were heterogeneous in nature. Synonymous techniques appeared under different names, and spelling conventions varied (e.g. 3D vs. 3-D). To address this, we developed a dedicated multi-stage pipeline, combining rule-based text processing, semantic embeddings, unsupervised clustering, and LLM-based reasoning to normalise, deduplicate, cluster, and classify these terms into a coherent two-level taxonomy.

The hierarchy is intentionally asymmetric and organised across three levels (Table 1). At the broadest level, categories are editorially grouped into seven thematic domains (L1): \emph{Statistics \& Modelling}, \emph{Space \& Territory}, \emph{3D \& Image Analysis}, \emph{Machine Learning \& AI}, \emph{Material Analysis}, \emph{Simulation \& Networks}, and \emph{Data \& Communication}. These domains serve only as a reading aid and reflect the major sub-disciplines of computational archaeology, but play no computational role in the analytical pipeline. The operative upper level (L2) consists of the 25 fixed, researcher-defined categories covering the full methodological landscape of the discipline, from classical statistical inference and Bayesian modelling to remote sensing, geometric morphometrics, agent-based simulation, and NLP (see Table S5.1 in the ESM for the full definitions, which are the ones supplied to the model in the assignment prompts). Each L2 category was provided with a descriptive definition and a list of representative methods, which were later used as anchoring context in the LLM prompts. Recognising that modern computational methods often cross-cut taxonomic boundaries (e.g. autoencoders operating simultaneously as dimensionality reduction tools and deep learning architectures \citep{Bowker1999sorting, Rivest2021article}), a single-label constraint was pragmatically enforced at the L2 level to maintain pipeline scalability. The lower level (L3) is instead fully data-driven: clusters are discovered from the corpus itself, without any prior enumeration, and labelled automatically by the LLM.

\begingroup\small
\begin{longtable}{p{0.22\linewidth}p{0.34\linewidth}p{0.38\linewidth}}
\caption{Overview of the predefined hierarchical taxonomy (L1 and L2 levels). Shows the seven broad thematic domains (L1) and the 25 operative categories (L2) with their representative methods, used as anchoring context in the LLM assignment pipeline (see Table S5.1 in the ESM for full operational definitions).}\label{tab:taxonomy} \\
\toprule
\textbf{Group (L1)} & \textbf{Category (L2)} & \textbf{Representative Methods} \\
\midrule
\endfirsthead
\toprule
\textbf{Group (L1)} & \textbf{Category (L2)} & \textbf{Representative Methods} \\
\midrule
\endhead
\midrule
\multicolumn{3}{r}{\footnotesize\textit{Continued on next page}} \\
\bottomrule
\endfoot
\bottomrule
\endlastfoot
\textbf{Statistics \& Modelling} & Univariate \& Classical Hypothesis Testing & t-test, Mann-Whitney U, Chi-square, ANOVA, Kruskal-Wallis, Kolmogorov-Smirnov \\
 & Multivariate Analysis \& Dimensionality Reduction & PCA, Correspondence Analysis, MDS, t-SNE, Procrustes Analysis, Factor Analysis \\
 & Regression \& Generalized Linear Models & OLS Regression, GLM, GAM, Logistic Regression, Mixture Modelling \\
 & Clustering \& Unsupervised Learning & k-means, Hierarchical Clustering, DBSCAN, HDBSCAN, Gaussian Mixture Models \\
 & Distance \& Similarity Metrics & Mahalanobis Distance, Nearest Neighbor Analysis, Jaccard Similarity, Edit Distance \\
 & Bayesian \& Probabilistic Inference & MCMC, Metropolis-Hastings, Gibbs Sampling, Hierarchical Bayesian Models, Bayesian Networks \\
 & Time Series \& Sequence Analysis & Autocorrelation, Fourier Analysis, Wavelet Transform, DTW, ARIMA \\
 & Chronological Modelling \& Dating & Bayesian Radiocarbon Calibration (OxCal), OSL Modelling, Dendrochronology, Seriation \\
\midrule
\textbf{Space \& Territory} & Spatial Statistics \& Point Pattern Analysis & Ripley's K, Kernel Density Estimation, Kriging, Moran's I, Geographically Weighted Regression \\
 & GIS \& Landscape Analysis & Viewshed Analysis, Least-Cost Path, Site Catchment Analysis, Digital Elevation Models \\
 & Remote Sensing \& Satellite Imagery & Multispectral Classification, SAR Analysis, NDVI, Change Detection, Image Segmentation \\
 & Geophysical Prospection & Ground-Penetrating Radar (GPR), Magnetometry, ERT, Seismic Refraction, EMI \\
\midrule
\textbf{3D \& Image Analysis} & Photogrammetry \& 3D Reconstruction & Structure-from-Motion (SfM), Multi-View Stereo (MVS), Terrestrial Laser Scanning, Point Cloud Registration \\
 & Geometric Morphometrics & Landmark-Based Morphometrics, Elliptic Fourier Analysis, Thin-Plate Splines, Procrustes Superimposition \\
 & Computer Vision \& Image Processing & Edge Detection, Texture Classification, Template Matching, Image Segmentation, Colour Analysis \\
\midrule
\textbf{Machine Learning \& AI} & Machine Learning \& Supervised Classification & SVM, Random Forests, Gradient Boosting, Decision Trees, k-Nearest Neighbours, Naive Bayes \\
 & Deep Learning \& Neural Networks & CNN, RNN, Transformers, Transfer Learning, Autoencoders, Generative Adversarial Networks \\
 & NLP \& Text Mining & Named Entity Recognition, Topic Modelling, Word Embeddings, LLMs, OCR, Corpus Analysis \\
\midrule
\textbf{Material Analysis} & Archaeometry \& Compositional Analysis & XRF/pXRF, Raman Spectroscopy, FTIR, Neutron Activation Analysis (NAA), ICP-MS \\
 & Isotope \& Bioarchaeological Analysis & Strontium Isotope Analysis, $\delta^{13}\text{C} / \delta^{15}\text{N}$, aDNA Analysis, Proteomics, Isotope Mixing Models \\
 & Geoarchaeology \& Sediment Analysis & Granulometry, Micromorphometry, Geochemical Profiling, Magnetic Susceptibility, Phytolith Analysis \\
\midrule
\textbf{Simulation \& Networks} & Agent-Based Modelling \& Simulation & Agent-Based Models (ABM), Cellular Automata, System Dynamics, Monte Carlo Simulation \\
 & Network Analysis \& Graph Theory & Social Network Analysis, Centrality Measures, Community Detection, Exchange Network Modelling \\
 & Ecological \& Palaeoenvironmental Modelling & Species Distribution Models (SDM), MaxEnt, Palynological Modelling, Climate Reconstruction \\
\midrule
\textbf{Data \& Communication} & Visualization \& Scientific Communication & Scientific Cartography, 3D Visualization, Virtual Reality (VR/AR), Interactive Web Maps, Dashboards \\
\bottomrule
\end{longtable}
\endgroup

The identified methods are processed as follows, producing 242 candidate L3 clusters of which 241 are retained for analysis (see Phase 4.5 below and Fig. 2). The full list, with member terms and corpus frequencies, is given in Table S5.2 of the ESM; the prompts governing each phase are reproduced in ESM, \S{}S4. 

\textbf{Phase 1: Text normalisation.} All raw method strings were first subjected to a normalisation pass to reduce superficial variation. This included: (i) systematic British-to-American spelling conversion (e.g. \emph{modelling} → \emph{modeling}, \emph{colour} → \emph{color}, \emph{visualisation} → \emph{visualization}) applied via regular-expression dictionary; (ii) removal of intra-word hyphens between alphabetic tokens; (iii) capitalisation standardisation, preserving all-uppercase acronyms while title-casing mixed-case terms; and (iv) stripping of common English plurals from multi-character words. These rules reduced the unique term vocabulary substantially before any similarity-based processing.

\textbf{Phase 2: Fuzzy deduplication.} Normalised terms were then grouped into equivalence clusters using fuzzy string matching via the \emph{\texttt{token\_sort\_ratio}} metric from the \emph{thefuzz} library (\url{https://github.com/seatgeek/thefuzz}), with a similarity threshold of 88 out of 100. This metric is order-insensitive, correctly collapsing permutations such as \emph{Principal Component Analysis} and \emph{Analysis Principal Component}. Within each fuzzy cluster, the most frequently occurring variant was selected as the canonical term representing the entire group. 

\textbf{Phase 3: Semantic embedding and L3 clustering.} Canonical terms were embedded into a high-dimensional semantic space using a Qwen3-Embedding-8B model \citep{Zhang2025qwen3embedding}. Each term was prefixed with a domain-specific instruction (`Represent this computational method for grouping by its specific algorithmic paradigm and technique family, ignoring application domain'\emph{)} following the instruction-tuned embedding paradigm to suppress topical noise and focus the representation on methodological rather than disciplinary similarity. The resulting embedding matrix was clustered using EVoC, a hierarchical density-based algorithm that produces a layered partition tree \citep{Bot2026persistent}. The finest-grained layer (layer 0) was retained as the L3 partition, with a minimum cluster size of 5 and a maximum of 12 hierarchy layers. Points classified as noise (label $-1$) were reassigned to the nearest cluster centroid via a k-Nearest neighbor classifier \citep{Goldberger2004neighbourhood}, ensuring full coverage. A final semantic merge pass iteratively fused pairs of L3 clusters whose centroids exhibited a cosine similarity of $\ge 0.98$, preventing near-duplicate clusters from reaching the labelling stage and reducing redundant LLM calls. 

\textbf{Phase 4: Automatic L3 labelling.} Each L3 cluster was labelled by the Qwen3.5-9B generative model via a structured chat-completion prompt. The model received the up-to-30 most frequent member terms of the cluster and was instructed to return a single short label (2--5 words). The prompt explicitly prohibited generic labels (e.g. \emph{Methods}, \emph{Techniques}, \emph{Analysis}) and compound labels joining two distinct families. Temperature was set to 0.1 to allow controlled lexical variation while maintaining coherence. This phase is common in topic modelling tasks, as it improves general interpretability\footnote{e.g. \url{https://maartengr.github.io/BERTopic/getting_started/representation/llm.html}).}.

\textbf{Phase 4.5: Garbage validation}. Density-based clustering over a vocabulary this heterogeneous inevitably leaves a residual cluster absorbing terms with no methodological content. Each labelled L3 cluster was therefore submitted to a single validation call instructed to flag a cluster only when a notable share of its members are not scientific methods (generic words, software product names, non-analytical activities), and to default to retaining the cluster whenever in doubt. Exactly one of the 242 clusters was flagged: its members include \emph{Buffer}, \emph{Comparison}, \emph{Control}, \emph{Choice} and \emph{Card Sorting}. Discarding it leaves the 241 L3 categories used throughout the analysis and removes only 6 of the 8,404 articles, since almost every affected article also reports at least one genuine method.

\textbf{Phase 5: L3-to-L2 assignment.} Each L3 cluster was then assigned to one of the 25 fixed L2 categories by the LLM. The full L2 catalogue including numeric index, label, and description for each category was embedded in the prompt, and the model was asked to return the index of the best-matching category. The prompt explicitly instructed the model to prioritise the algorithmic and statistical nature of the methods over their archaeological application domain, avoiding spurious assignments driven by disciplinary context rather than methodological type. 

\textbf{Phase 6: Adversarial review.} To correct residual misassignments, each L3-to-L2 mapping was subjected to an independent triple-vote review: the LLM was called three times with a conservative auditing prompt that defaulted to confirming the current assignment unless the category was deemed completely unrelated to the cluster's methodology. A reassignment was applied only when all three votes unanimously agreed on the same alternative category. This majority-unanimity design minimises false corrections: a single dissenting vote was sufficient to preserve the original assignment, making the review pass a conservative quality-control filter rather than a re-classification \citep{Bougie2024generative}.

\begin{figure}[htbp]
\centering
\includegraphics[width=0.85\linewidth]{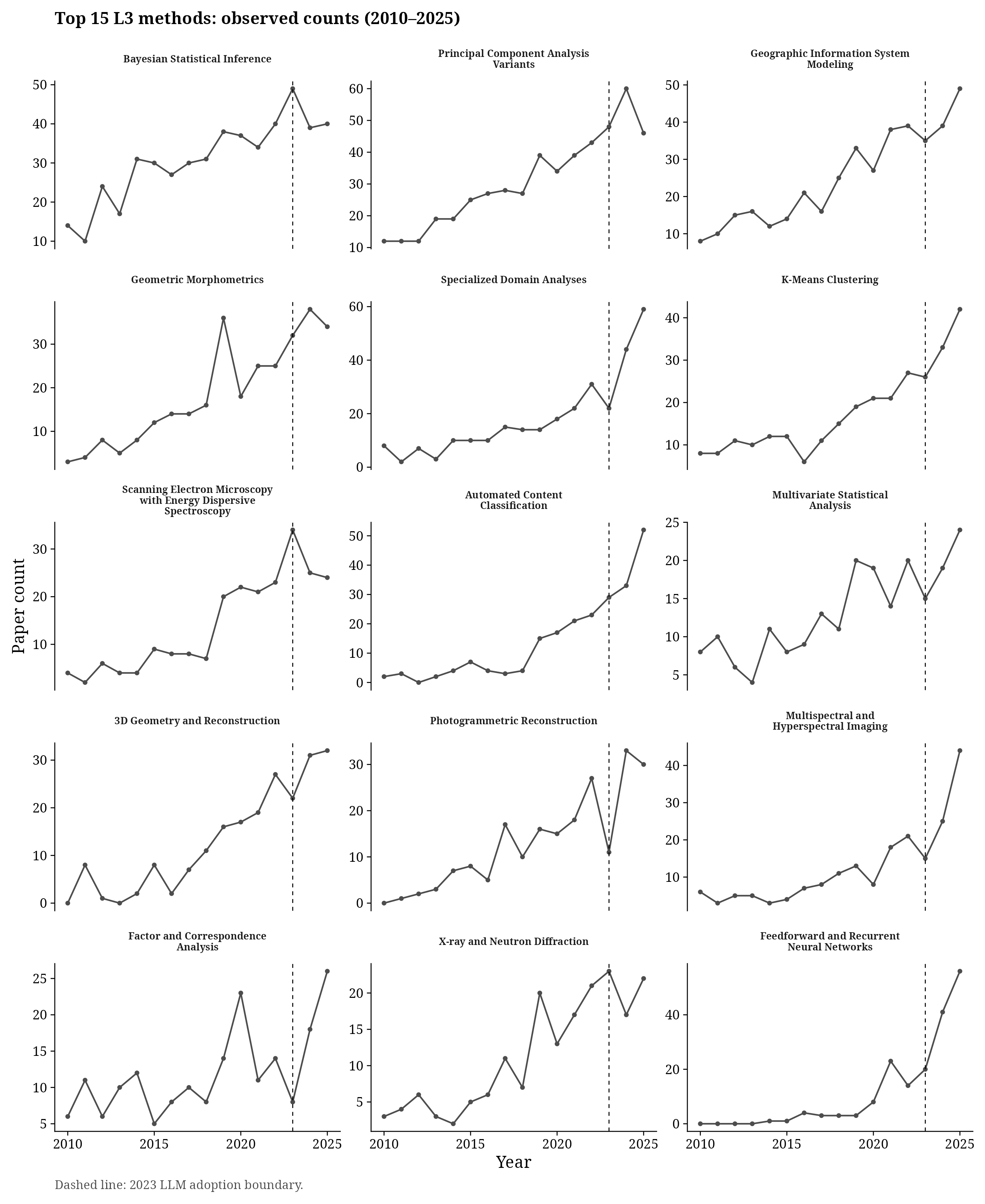}
\caption{Raw counts of L3 methods extracted from the bibliometric stream (2010--2025). The figure shows the top 15 methods. The dashed vertical line marks the 2023 post-LLM boundary.}
\label{fig:raw_counts}
\end{figure}

\subsection{Analysis of the Bibliometric Stream}\label{sec:analysis_biblio} 

The resulting corpus of L3 technique counts within each L2 sub-discipline was modelled using a Bayesian hierarchical Dirichlet-Multinomial regression, fit in \emph{Stan} via the R (Version 4.6) package \emph{cmdstanr} \citep{Gabry2025cmdstanr}\footnote{The analyses were performed on a Ubuntu 24.04.4 running on a Dell Inc. PowerEdge R650xs with 128GiB RAM and 96 Intel(R) Xeon(R) Gold 6336Y CPU @ 2.40GHz processors each with 36864 KB cache.}. The Dirichlet-Multinomial model considers method counts in a compositional way: within any sub-discipline, an increase in one technique's share must be offset by a decrease elsewhere in the composition.  
The aim is to test if certain computational techniques are more frequent in archaeology after the wider availability of LLMs, and if we are able to capture this change in trajectory. Other models treating techniques as independent observations (e.g. a Poisson regression) cannot distinguish between field-wide growth and a specific technique becoming more popular at the expense of another. Our study focuses on the latter case.   
For each technique, the model separates three elements we are interested in measuring: (i) where a technique lies on average (the baseline log-share), (ii) its pre-existing trend before LLMs became common, and (iii) any additional shift after 2023 over and above that trend. The last component is the key figure in our regression. If its scale across techniques is credibly above zero, this indicates that methods diverged unevenly after 2023 and significantly from their prior trajectories. If, instead, the scale is near zero, it would mean that LLMs did not impact the methods of the newly published set of papers.   
Both the trend and the post-2023 shift are modelled hierarchically, as exchangeable across techniques within each sub-discipline. This implies that techniques within a sub-discipline share a common distribution, and the method has the further advantage of pulling extreme estimates back towards the group mean, a standard regularisation practice. 

\subsection{LLM Experiment Stream}\label{sec:experiment_stream}

To test whether LLMs act as a structural driver of methodological convergence in the discipline, a controlled simulation experiment was designed in which a local LLM was asked to recommend computational methods for a set of standardised archaeological research problems (Table 2). These outputs are referred to as `L4 methods' hereafter, as they are more granular than the L3 clusters of the taxonomy. These L4 methods were then mapped back onto the L3 taxonomy and compared with the estimated empirical frequency shifts from the Bayesian model. The core hypothesis, which we refer to as the \emph{mean-collapse hypothesis}, is that large language models trained on an aggregate of scientific literature will systematically favour the most statistically canonical methods when asked to provide recommendations. This bias will manifest as a positive association between the prevalence of a method in pre-2023 literature and its frequency of recommendation, particularly under minimal methodological guidance. In such cases, the model has the greatest freedom to default to the prevailing methods. Twenty-eight research questions were constructed to cover the full thematic range of the Scopus corpus, spanning all major sub-disciplines of archaeology across periods and regions (Table S6.1 in the ESM, which also lists the nine vague descriptors in Table S6.2 and reproduces the three profile prompts in \S{}S6.4). The questions were deliberately formulated in substantive, methodologically neutral terms as they describe an archaeological problem without implying any computational solution to avoid lexical priming of the model's output, which would be particularly damaging for the novice profile where the question is the only input. Nine broad methodological descriptors were defined to represent the kind of informal prior knowledge a researcher with partial computational training might hold: from `\emph{some kind of statistical or quantitative analysis}' to `\emph{some kind of simulation or agent-based modelling}'. These are purposefully imprecise and do not map one-to-one onto any L2 category, requiring the model to bridge from an informal description to a specific recommendation. The experiment simulates three researcher profiles that differ along a single dimension: the degree of methodological prior knowledge contributed to the prompt. Everything else (i.e. the model, the system prompt framing, and the output format instruction) is held constant across profiles.

\begin{table}[htbp]
\centering
\small
\caption{User profile formulations and representative prompts evaluated against the benchmark archaeological question: `How do I analyse the distribution and organisation of human settlements across a territory over the long term?' (identical across all profiles).}
\label{tab:profiles}
\begin{tabular}{p{0.18\linewidth}p{0.38\linewidth}p{0.38\linewidth}}
\toprule
\textbf{Profile} & \textbf{Description} & \textbf{Example Prompt} \\
\midrule
\textbf{Profile A (Expert)} & Profile A (Expert) provides a specific L2 category as prior knowledge and asks the model for concrete tools, algorithms, or variants within that family applicable to the given problem. This profile constrains the output space and is expected to produce the most diverse and technically detailed recommendations. & `\emph{I am working on: How do I analyse the distribution and organisation of human settlements across a territory over the long term? I already know I want to apply Spatial Analysis and Modelling to my analysis. Which specific tools, algorithms, or variants of this method would you recommend, and how would you apply them concretely to this problem?}' \\
\midrule
\textbf{Profile B (Intermediate)} & Profile B (Intermediate) provides only a vague methodological family, leaving both the choice of specific method family and the concrete technique to the model. This profile introduces partial constraints. & `\emph{I am working on: How do I analyse the distribution and organisation of human settlements across a territory over the long term? I have a rough idea that I need some kind of spatial or geographic approach, but I do not know which specific method to choose. What would you recommend?}' \\
\midrule
\textbf{Profile C (Novice)} & Profile C (Novice) provides only the archaeological problem, with no methodological guidance whatsoever. The model has full freedom over both method family and specific technique. This profile is expected to produce the strongest mean-collapse signal, as no prior knowledge is available to divert the model away from its default distributional tendencies. & `\emph{I am working on: How do I analyse the distribution and organisation of human settlements across a territory over the long term? I have no specific computational background. Which digital methods could I use to address this research problem?}' \\
\bottomrule
\end{tabular}
\end{table}

For each of the 252 iterations (28 questions $\times$ 9 vague families), a single triplet (i.e. one research question, one L2 category, and one vague descriptor) was selected and sent to all three profiles simultaneously. Thus, the three profiles in a given iteration face identical archaeological content and differ only in the methodological framing of the prompt. This within-iteration design enables direct comparisons between profiles while controlling for variability at the question level. The total number of generation calls was 252 $\times$ 3 = 756, plus three mapping calls per extracted L4 method. The experiment was run with two local LLMs. The primary model was Qwen3.5-9B \citep{Yang2025qwen3} and the secondary model was Gemma 4 E4B \citep{Abd2026gemma4, Gemma2026welcome},  run with an identical pipeline, identical prompts, and the same sampling design. Using two models from different training corpora allows us to distinguish structural properties of LLM recommendation behaviour in general from artefacts specific to one model's training distribution. Temperature was set to 0.1 across all calls. Each generation response was parsed line by line into a list of (method name, justification) pairs. The justification sentence was retained alongside the method name as a qualitative trace for post-hoc inspection of the model's reasoning. Responses that yielded no parseable items were stored as a single unparsed entry to avoid silent data loss. Each extracted L4 method was mapped onto the L3 taxonomy by a second LLM call, run three independent times per item. The mapping prompt provided the full list of L3 labels and asked the model to return the single most appropriate label exactly as it appears in the list, with no additional output. The three runs were compared: if all three agreed, the item was flagged as \emph{consistent}; if they disagreed, the majority-vote result was retained but the item was flagged for manual review. Label extraction from model responses followed a priority cascade, exact string match, numeric code prefix match (e.g. L3-097), longest substring match against the descriptive part of any label with raw response text retained as a fallback and automatically flagged as inconsistent. Across all mapping attempts, the raw method strings produced by the two models, 2,904 (Qwen) and 1,746 (Gemma) distinct strings after normalising for case and whitespace, collapsed almost entirely onto the existing L3 vocabulary: 194 and 167 of the 242 candidate labels respectively, or 193 and 166 of the 241 analytically retained labels, since both models mapped a small number of items into the cluster discarded at Phase 4.5\footnote{One item, in the Gemma run, could not be mapped to a valid L3 label: the model returned the string L3-27, a malformed rendering of an L3 code, since the taxonomy uses three-digit codes.}. Unlike the concentration and correlation analyses below, which are restricted to majority-consistent mappings across the three independent runs, this check uses all mapping attempts. The reason for this is that detecting out-of-taxonomy items does not require cross-run agreement. Across three independent mapping runs per item, the two models returned unanimous labels for 96.2\% and 96.5\% of items; the remainder were resolved by majority vote and flagged for review. No generation response failed to parse (ESM, \S{}S6.6).

\subsubsection{Concentration Analysis}\label{sec:concentration}

Before testing the main question (whether the introduction and popularisation of LLMs caused a collapse in the computational archaeology literature) we quantified how narrow the recommendation profile of each local LLM (Qwen or Gemma) is in relation to the published literature. To do so, the effective number of methods was computed for each distribution using the inverse Simpson index. This index converts a frequency distribution into the number of equally frequent methods that would produce the same degree of concentration: a distribution dominated by a few techniques scores low, whereas an even spread across many techniques scores high.   
A Bayesian model was fitted to the observed frequencies over the full L3 taxonomy for each source (pre-2023 and post-2023 literature) and each LLM profile combination. Before any data were observed, the model assumed that every method was equally likely (using a uniform prior). The observed counts from the local LLMs then update this assumption, assigning high estimated proportions to methods that are suggested many times and low proportions to those rarely suggested. The result is a posterior distribution over all possible proportion vectors for each condition. As this model has a simpler structure than the hierarchical regression model described in \S\ref{sec:analysis_biblio}, with a single set of proportions rather than groups, years and trend parameters, the posterior distribution can be computed exactly without the need for iterative MCMC sampling (see the ESM, \S{}S9 for technical details). We drew 4,000 samples from each posterior and later computed the inverse Simpson index for each sample. This generated a full distribution of values for the effective number of methods. The aim was to enable comparison of the credible intervals across all conditions, answering questions such as `Is the novice profile more concentrated than the expert?' and `Are the local LLM suggestions credibly narrower than the post-2023 literature?'\footnote{The recommendations within a single LLM response and the methods within a single paper are not independent of each other, but the model treats every count as independent. This results in the credible intervals being estimated as somewhat too narrow. Cluster bootstrapping (resampling of entire responses and papers) confirms that the effect is minimal and inconsistent: the intervals widen by up to ~2.9 times in the most clustered condition, by an insignificant amount elsewhere, and never by a sufficient amount to close the LLM--literature gap reported in the \S{}4 section.}. Assuming that the mean-collapse mechanism operates as hypothesised, the LLM distributions should be substantially more concentrated than those in the literature. The largest concentration gap should be observed for the novice profile, where the model receives no methodological constraints. The following section investigates what drives LLM method recommendations.

\subsubsection{What Drives LLM Method Recommendations?}\label{sec:what_drives}

The aforementioned LLM experiment and concentration analysis were designed to test the validity of our \emph{mean-collapse hypothesis}. To do so, we modelled the number of times that each L3 technique was recommended by the LLM as a function of two predictors using a Bayesian negative binomial regression, a standard choice for overdispersed count data, as it was reasonable to expect that a few methods may have been recommended many times, while other methods received just a few (see ESM, \S{}S10). \begin{equation}
n_{\text{rec}, i} \sim \text{NB}_2\left(\exp(\alpha + \beta_{\text{pre}} \log(1 + n_{\text{pre}, i}) + \beta_\gamma \gamma_i),\, \phi\right)
\label{eq:nb_model}
\end{equation}

where $n_{\text{pre}, i}$ is the pre-2023 literature count for method $i$ and $\gamma_i$ is the signed post-2023 excess. The first predictor is pre-2023 prevalence ($\beta_{\text{pre}}$): a model that recommends methods in proportion to their frequency in the archaeological literature prior to the introduction of LLMs will reproduce the aggregate of its training distribution and systematically favour what the field has relied upon since 2010. This predictor also represents the direct operationalisation of the mean-collapse mechanism. The second predictor, post-2023 share gain ($\beta_\gamma$), determines whether additional variance is explained by recent disciplinary movement beyond the pre-2023 prevalence. Under the mean-collapse mechanism, this predictor is not expected to produce a positive outcome.  
If the \emph{mean-collapse effect} is real, the pre-2023 prevalence coefficient should exhibit a consistent pattern across profiles. A novice prompt gives the LLM full control over its response, and the pull towards more canonical methods should be the most visible here. In contrast, following our experimental design (ESM, \S{}S7), an expert prompt constrains the output space, attenuating the signal.  
We fitted the model eight times, once for each combination of two LLMs (Qwen3 and Gemma) and four profiles (novice, intermediate, expert and all pooled together). As already mentioned, we designed the experiment with two local LLMs to test if the pattern holds across architectures trained on different corpora. In this case, the effect would be more likely to be a structural feature of how LLMs process archaeological and methodological questions than an artefact of the training history of any single model. 

\section{Results}\label{sec:results}

The analysis of the archaeological literary corpus shows an increased application of computational methods in the discipline. Among these, specific methods gained particular traction since 2010 (i.e. the lowest boundary of our study), which is already observable from the raw counts of the L3 methods extracted from the bibliometric stream (Fig. 2).  
In order to distinguish compositional shifts after the introduction of LLMs from pre-existing disciplinary trends, we employed a Bayesian hierarchical Dirichlet-Multinomial regression, which was fitted to the full computational archaeology corpus. The model converged well with $\hat{R}$ < 1.01, bulk effective sample sizes above 640, with zero divergent transitions\footnote{Further technical details and diagnostics can be found in the ESM, \S{}S7.}.   
The regression was designed to answer a key question: had there been an unusual shift in computational archaeology literature since LLMs were introduced? Although the number of linguistic markers clearly associated with LLMs is on the rise and already visible in published literature (Fig. 1), it is difficult to gain a clear understanding of vibe coding practices using data from just three years of publications. Firstly, the model confirms that the field exhibits compositional regularity, with stable and predictable proportions of methods within sub-disciplines from year to year. This regularity provides a favourable signal-to-noise ratio for detecting genuine structural change. Against this stable backdrop, the post-2023 variation in methods ($\sigma_\gamma$) is small, but credibly above zero (90\% CI $\approx$ 0.01--0.22), whereas the pre-LLM baseline variation ($\sigma_\beta$) is approximately 2.35 times greater (see Fig. 3). In plain terms, while some methods did shift after 2023, the magnitude of this shift is modest compared to the long-term heterogeneity that the field has always displayed. At the level of individual methods, no single L3 technique produced a credible estimate of change at the 90\% threshold (none of the 241 methods did so). Thereby, although the signal is detectable in aggregate, it cannot be localised to any specific method.

\begin{figure}[htbp]
\centering
\includegraphics[width=0.85\linewidth]{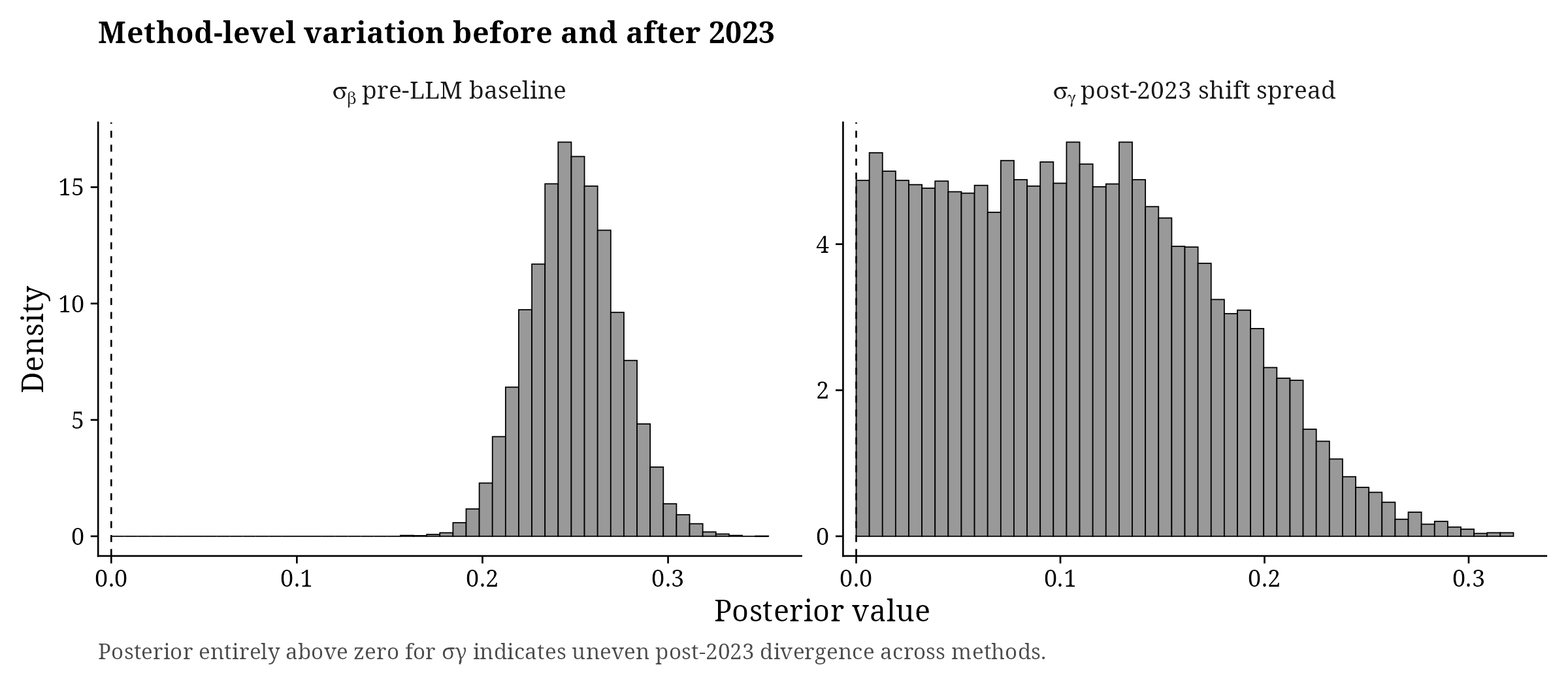}
\caption{Posterior distributions for the $\sigma$ parameters, showing the baseline trend ($\sigma_\beta$) vis-\`a-vis the substantial heterogeneity after the large language models shift ($\sigma_\gamma$).}
\label{fig:sigma_posteriors}
\end{figure}

Indeed, when we used the Inverse Simpson index to measure the effective number of L3 methods at field level, we observed an increase from 87.6 (90\% CI: 84.6--90.6) to 111.2 (90\% CI: 107.7--114.7) after 2023. Rather than converging towards fewer techniques, it seems that computational archaeology papers have become more diverse since the introduction of LLMs. This counterintuitive finding is at odds with our original hypothesis, indicating that LLMs have not, at least so far, produced the methodological convergence we anticipated.  
If the analysis of the bibliometric stream was not entirely conclusive, the results of the concentration analysis are less ambiguous. When two different LLMs were given deliberately neutral instructions, they recommended approximately one third of the effective number of methods observed in the literature (Table 3). In contrast to the effective number of L3 methods, both models fell short: Qwen3 produced an overall total of 31.6 (90\% CI: 30.2--33.0) effective methods, while Gemma produced 28.8 (90\% CI: 27.6--30.0). The gap is credible across all comparisons, with Qwen3 being marginally more diverse than Gemma (Fig. 4).

\begin{figure}[htbp]
\centering
\includegraphics[width=0.85\linewidth]{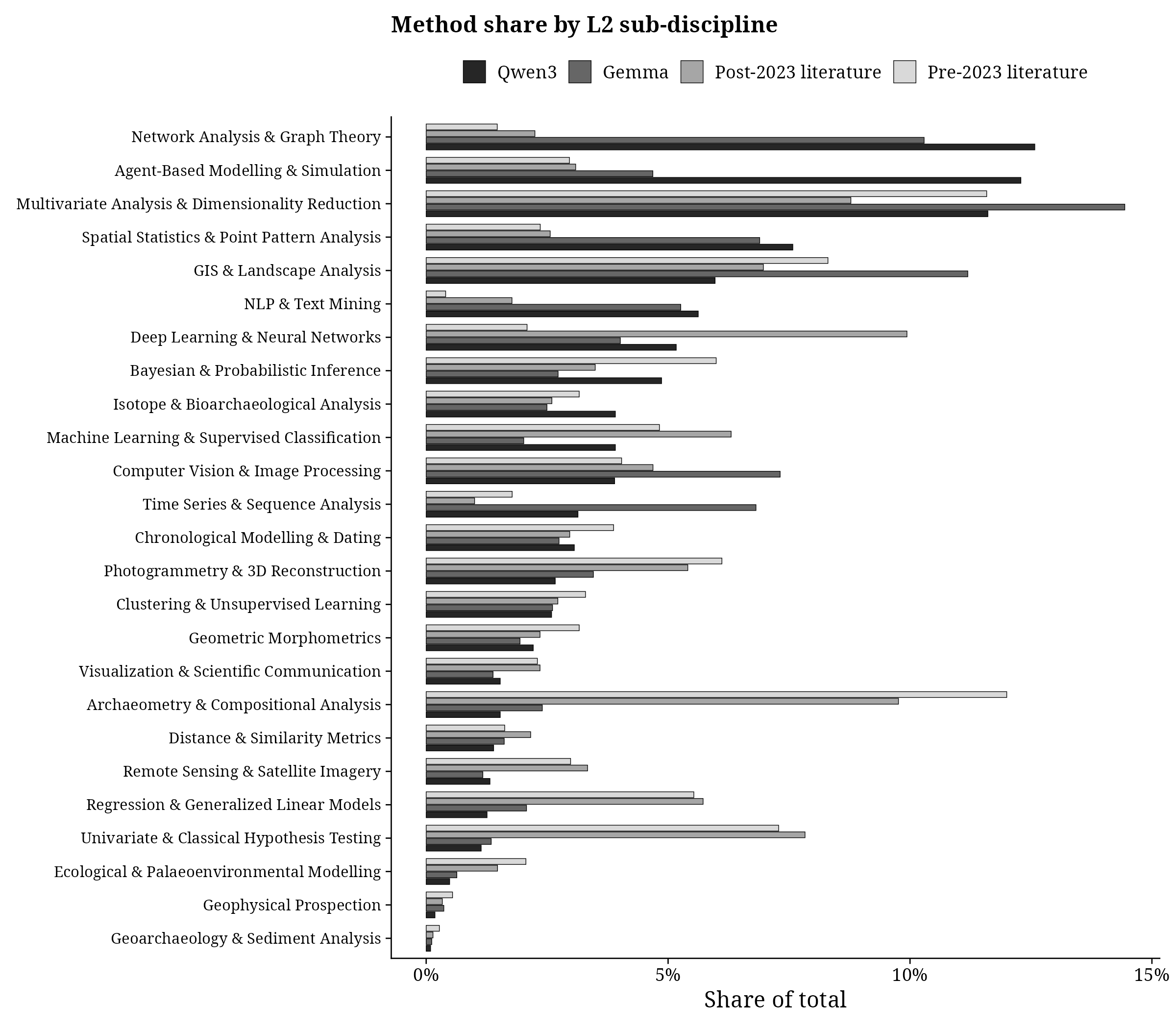}
\caption{Method shares across the 24 L2 sub-disciplines, comparing Qwen3 and Gemma recommendations with pre- and post-2023 literature.}
\label{fig:l2_shares}
\end{figure}

The extent of this concentration is shaped by the strength of the methodological guidance in the prompts. For novices with no prior knowledge (i.e. generic prompts), the effective number of methods narrows to around 21 for both models, which is roughly a quarter of the pre-LLM literature. Interestingly, this figure increases substantially when a methodological domain is specified: expert profiles reach 60.7 (Qwen3, 90\% CI: 56.8, 64.7) and 46.8 (Gemma, 90\% CI: 43.4--50.2). While this gradient remains consistent across both models, none of the profiles matches the breadth of methods in the published articles. This result can be understood partly in terms of our prompt design: the expert profile instructs the model to work within a specific methodological framework, thereby limiting the scope for default responses. However, as the two models are trained on different corpora, this result may be a structural feature rather than simply an incidental outcome derived from our prompt design \citep{Jiang2025evaluating}.

\begin{table}[htbp]
\centering
\small
\caption{Effective number of L3 methods (inverse Simpson index) by source, with the Qwen3 and Gemma sets broken down by researcher profile. Values are posterior medians with 90\% credible intervals from the Dirichlet conjugate model (\S\ref{sec:concentration}).}
\label{tab:effective_methods}
\begin{tabular}{llcc}
\toprule
\textbf{Source} & \textbf{Profile} & \textbf{Median} & \textbf{90\% CI} \\
\midrule
Literature pre-2023  & ---          & 87.6  & [84.6, 90.6]   \\
Literature post-2023 & ---          & 111.2 & [107.7, 114.7] \\
\midrule
Qwen3                & Novice       & 20.9  & [19.5, 22.4]   \\
Qwen3                & Intermediate & 30.5  & [28.3, 33.0]   \\
Qwen3                & Expert       & 60.7  & [56.8, 64.7]   \\
Qwen3                & Overall      & 31.6  & [30.2, 33.0]   \\
\midrule
Gemma                & Novice       & 20.5  & [19.0, 21.9]   \\
Gemma                & Intermediate & 29.5  & [27.3, 31.7]   \\
Gemma                & Expert       & 46.8  & [43.4, 50.2]   \\
Gemma                & Overall      & 28.8  & [27.6, 30.0]   \\
\bottomrule
\end{tabular}
\end{table}

The methods that each model relies on most heavily are largely the same, with network analysis and modelling, discrete and agent-based simulation, and GIS modelling heading both lists (see Fig. 5). Several of these methods have plateaued or declined in the literature published since 2023, which suggests that the frequency with which they are recommended tracks past prevalence rather than recent momentum.

\begin{figure}[htbp]
\centering
\includegraphics[width=0.85\linewidth]{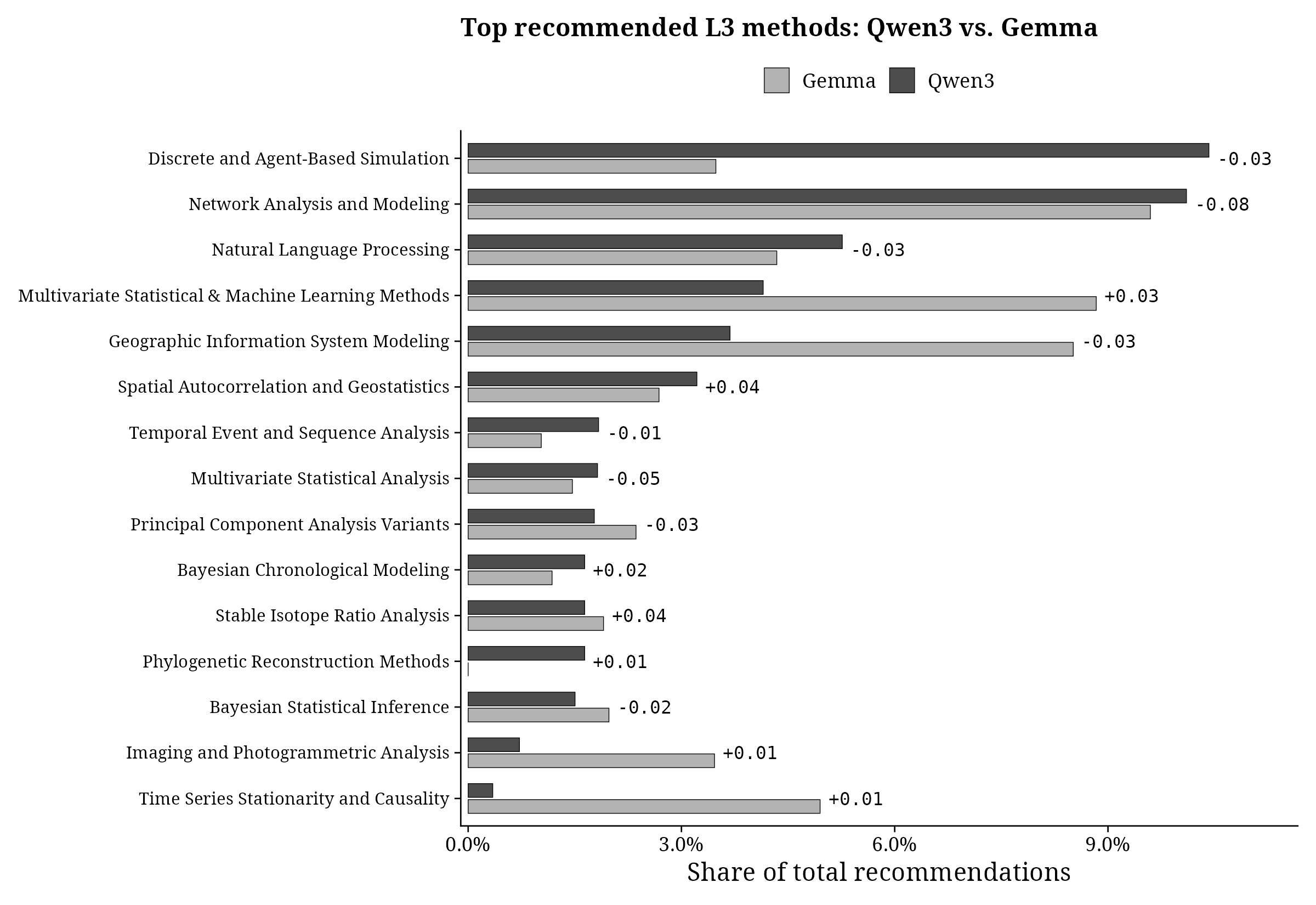}
\caption{Top ten recommended L3 methods per model, as a share of that model's recommendations. The label beside each bar is the method's post-2023 literature trajectory ($\gamma$, \S\ref{sec:analysis_biblio}); positive means gaining share, negative means declining relative to the pre-2023 trend.}
\label{fig:top10_methods}
\end{figure}

Finally, the Negative Binomial regression gives us the ability to distinguish between two reasons why a method might be frequently recommended by an LLM: (i) if it was prevalent in pre-2023 archaeological literature; or (ii) its usage increased after LLMs were introduced. The former is positively correlated in both models and profile combinations, with a magnitude that follows the gradient already observed in the concentration analysis: the largest magnitude is observed for the novice profile (Qwen3: 1.039, 90\% CI: 0.810--1.275; Gemma: 0.675, 0.378--0.984), and the smallest magnitude is observed for the expert profile (Qwen3: 0.419, 0.256--0.578; Gemma: 0.511, 0.338--0.685). Here, a more comprehensive problem formulation leaves less scope for the default use of standard methods. Conversely, the post-2023 term has no detectable signal (with 90\% confidence intervals ranging from -1.5 to +1.8). While the positive pre-2023 association directly operationalises the mean-collapse mechanism, the null post-2023 term is non-informative. This is to be expected, as the bibliometric analysis showed that no method shifted credibly after 2023.

\begin{figure}[htbp]
\centering
\includegraphics[width=0.85\linewidth]{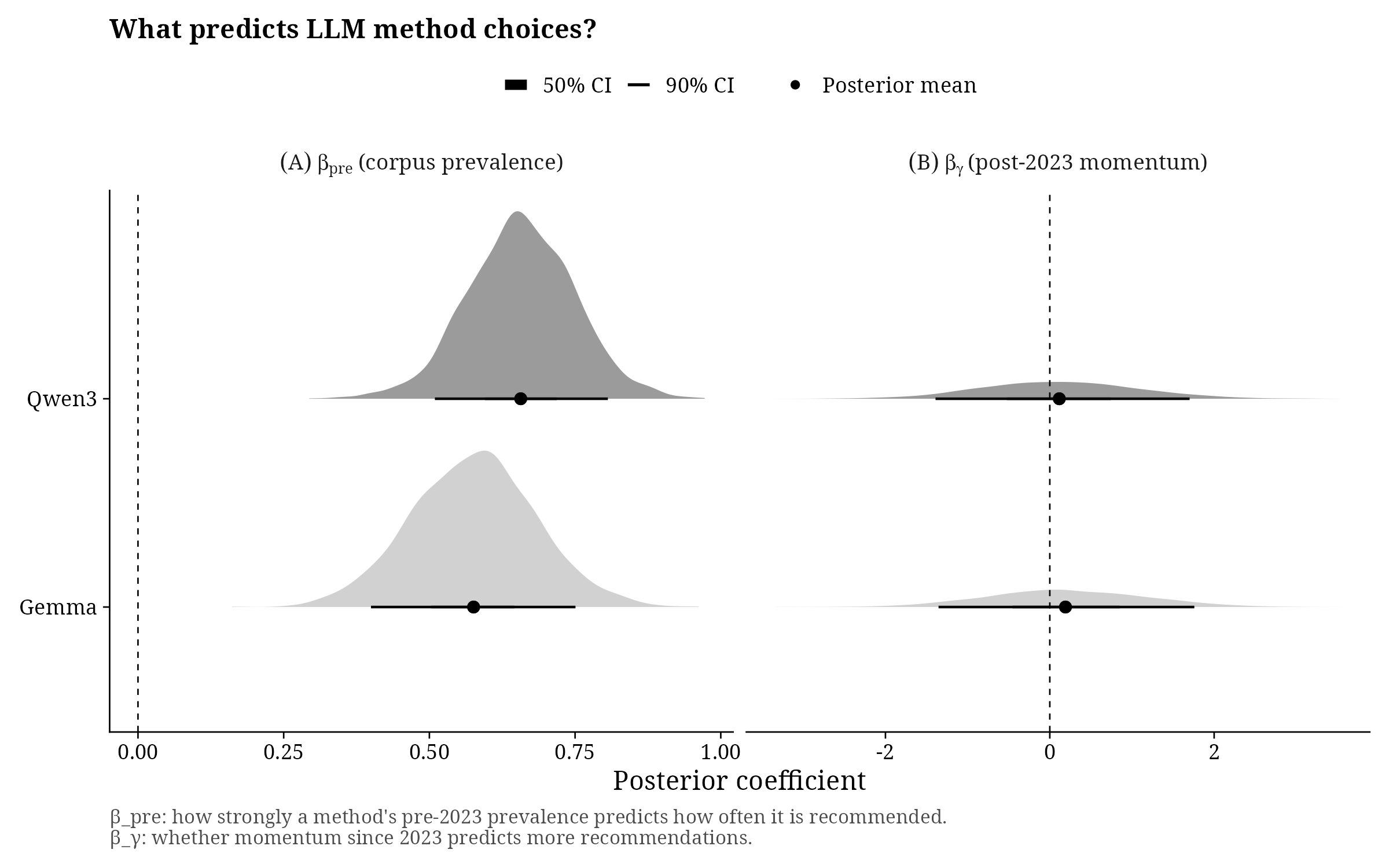}
\caption{Posterior distributions from a two-predictor count regression testing what drives LLM method recommendations. $\beta_{\text{pre}}$ measures how strongly a method's pre-2023 prevalence in the literature predicts how often it is recommended; $\beta_\gamma$ measures whether methods that gained momentum after 2023 are recommended more often. Point: posterior mean; thick bar: 50\% credible interval; thin bar: 90\% credible interval; dashed line: zero.}
\label{fig:regression_posteriors}
\end{figure}

\section{Discussion}\label{sec:discussion}

This research began with a structural question: whether the practice of vibe coding, now widespread across the software development sector and increasingly in academia, has already produced a mean collapse in archaeological literature. At first glance, our results seem to be conflicting. On the one hand, our prompting experiments established that this mechanism could potentially apply also to quantitative archaeology. Across different models (trained on different corpora) and prompting strategies, LLMs consistently favoured a narrower subset of computational and statistical methods than those present in the published literature, particularly when the prompts were naive or underspecified. Overall, these recommendations range from a third of the effective methods extracted from archaeological papers, to as little as a quarter under novice prompting. This pattern mirrors experimental findings in other domains. \citet{Doshi2024generative} showed that writers using AI-generated story ideas produced individually more novel work while collectively converging on markedly more similar outputs, and \citet{Anderson2024homogenization} reported the same divergence between individual- and group-level diversity in AI-assisted ideation. Under these settings, both LLMs reproduce the pre-2023 literary canon in direct proportion to its prior frequency.   
On the other hand, the bibliometric stream that we analysed instead does not yet demonstrate methodological contraction. In fact, the effective number of computational methods appears to have increased after 2023 (from 87.6 to 111.2), with no significant reshuffling among the individual techniques. While both LLMs suggest that a mechanism capable of homogenising archaeological literature appears to exist, its effects are not yet evident.   
Several factors might explain this inconsistency, which we anticipated as one of the two possible outcomes of our analysis, in Section~\ref{sec:aims}. First, the bibliometric analysis and the prompting experiment are not measuring exactly the same stage of the research process. The bibliometric stream records the methods ultimately reported in published abstracts, whereas the prompting experiment concerns how researchers could be guided into methodological decisions. The literature indeed showed a clear increase in linguistic markers commonly associated with LLM-assisted writing after 2023. However, we should be careful not to confuse writing assistance with methodological assistance, the latter of which is more difficult to measure. If a method is suggested by an LLM, there would not be a detectable linguistic trace in the final publication: in other words, if we can infer that archaeologists use these tools as a writing aid, we cannot definitely state that they are thinking their methods through LLMs.  
A second explanation concerns who is currently adopting these tools. The current adopters are unlikely a representative sample of the discipline. More plausibly, AI-assisted code is being used by researchers with pre-existing programming expertise, who are able to formulate targeted prompts and evaluate the generated code more critically. Thinking in terms of our experiment, these would match the `expert' profile more closely than the novice one. If this is the case, the mean-collapse effect would be relatively weak as the expert profile, while still concentrating below the literature (60.7 and 46.8 effective methods for the two models), is markedly broader than the novice profile (around 21). Archaeologists with limited programming experience might take longer to adopt these tools, potentially delaying the effects of methodological convergence in the literature.  
We cannot rule out the possibility of a third force acting in the opposite direction. Vibe coding may act as a diversifying force rather than reducing methodological diversity. Technical barriers associated with statistical programming and software development are undoubtedly lower now, and LLMs increasingly allow researchers to perform analyses that would previously have required collaboration with specialists or extensive programming expertise. The increase in methodological diversity observed after 2023 may reflect the fact that researchers can now implement a wider range of computational approaches more quickly and have increased access to resources that were previously unavailable to non-specialist researchers.  
This should, of course, be read as an interpretation rather than a definitive result. Our study cannot determine whether this diversification is caused by AI-assisted coding or by broader disciplinary developments, such as field growth or increasing interdisciplinarity. The two explanations are not mutually exclusive,  and the inverse Simpson increase is consistent with all of these.  
Taking all these factors into account, we can conclude that two opposing processes are occurring simultaneously. The first process expands the range of methods and tools available for archaeological research. It is probably used by users with a good understanding of their methods, who use these tools to remove resource constraints rather than outsourcing their judgement. The second process, which we observed from our prompting experiments, is capable of standardising the methods within the discipline and is, for the time being, still latent.   
The present diversification is what one would expect when the first pressure dominates. The structural risk lies in the composition of the second: as the use of LLM assistance becomes more widespread among those entering the discipline, the diversifying pull weakens and the homogenising pull grows.  
The net sign we observe in 2025 reflects \emph{who} is using these tools today, not \emph{what} the mechanism will produce as the balance between the two processes shifts.   
Following our analysis, the more imminent question is how quickly the transition to vibe coding is taking place. Explicit disclosure of AI use for code generation remains very rare in journal articles, even though anecdotal evidence and discussions among colleagues suggests the practice is far from uncommon. A longitudinal analysis in the future might be better placed to identify when this transition becomes visible in the published literature. For the time being, AI-assisted coding remains largely unregulated, and its uncritical use has all the potential to produce long-term consequences.

We now move from the quantitative findings to address the second objective of this paper:  the epistemological implications of generative AI within archaeology. As a discipline, archaeology has continually borrowed methods and technologies from neighbouring fields. With AI innovations constantly dominating the headlines, research institutions are increasingly providing access to generative coding tools\footnote{e.g. \url{https://oerc.ox.ac.uk/ai-centre/generative-ai-tools}; \url{https://help.uis.cam.ac.uk/news/google-gemini-and-notebooklm-available-universitys-google-workspace};  \url{https://www.uni-goettingen.de/en/686446.html}}. We then believe that the question is no longer whether archaeologists will adopt GenAI, but when, and on what terms.

As early career researchers, we are exposed to the persuasive advantages of AI-assisted coding. These tools can automate routine analytical tasks, some of which constitute the majority of an early-career computational archaeologist’s workload. More importantly, however, they reduce technical barriers, enabling analysis that would previously have been impossible without advanced programming expertise. This is particularly relevant in a  discipline where formal training in statistics and programming is often limited to a handful of universities in the Anglosphere and Northern Europe, and where the introductory literature \citep{Carlson2017quantitative, Drennan2009statistics, Shennan1997quantifying} is rarely translated into other languages. The democratising promise of AI-assisted coding is however conditioned by several factors\footnote{We set out some of these conditions in Section \S{}1.1.}: the cost asymmetry between consumer tiers and research-grade access \citep{StokelWalker2026ai}, data-retention policies that obstacle ethical research, and the `Matthew effect' of cumulative advantage \citep{Merton1968matthew} by which privileged institutions are likely to consolidate their advantages, as access to advanced computational tools in education directly translates into higher research output and visibility. In fact, the cost of individual subscriptions is virtually always on the rise, which penalises students and researchers from lower-income households. 

What our results add to this account is an empirical edge. The novice profile, which is the least able to constrain the LLM output, is the most exposed to the model’s defaults, and likely coincides with the same researchers that are least able to afford the tools that would make those defaults more navigable. Access inequality and methodological homogenisation are not distinct problems but a single issue.

Another specific disciplinary question involves \emph{who} is authorised to develop computational tools in the first place. Vibe coding effectively lowers the entry barrier for humanists without formal training in statistics or programming, thereby enabling them to produce functioning analytical pipelines. This could be viewed as either a long-overdue diversification of who gets to engage in computational archaeology, or as a form of de-skilling, whereby practitioners acquire what \citet{Collins2007rethinking} term \emph{interactional} rather than \emph{contributory} expertise. Interactional expertise is the ability to talk through and carry out an analysis without having the necessary skills to construct, audit or troubleshoot it independently. This tension is evident not only in archaeology, but also across the entire scientific landscape. \citet{Matsui2026return} shows that the long-term decline in solo-authored publications has been reversed since the release of ChatGPT, most strongly in fields where a coauthor's technical contribution is most easily replaced by AI, and that these solo papers tend to narrow in scope and shift toward computational topics, mirroring the mean-collapse dynamic we test here at the level of individual researchers. 

Our results seem to address this tension, but we have not yet been able to fully resolve it. The novice profile in our experiments captures this scenario in its simplest form: users who can obtain a functioning analytical pipeline from an LLM, but whose ability to judge whether its recommendations are appropriate is limited by design. Whether this expansion of the discipline’s practitioner base is a positive development or instead leads to a proliferation of unaudited, black-boxed analyses may depend less on the tools themselves than on how archaeological training and peer review adapt. Producing code is no longer necessarily the same as understanding it.

While computer science departments are increasingly integrating generative AI into their curricula and allowing students to use these tools to varying degrees in coursework \citep{KohenVacs2025integrating, Nathaniel2025literature}, other disciplines appear to be adopting them more at a slower pace. The use of the word `slow' here is deliberate. Concerns around \emph{slow science} \citep{Stengers2018another} and \emph{slow archaeology} \citep{Caraher2019slow} predate the `AI turn' in the digital humanities, often being situated within broader critiques of acceleration and productivity in the production of academic knowledge (see also \citealt{Marila2019slow} and \citealt{Caraher2019slow}). However, in the current \emph{publish or perish} ecosystem, is a `Slow AI' approach (i.e. intentionally decelerating technological integration and reintroducing methodological friction) practically feasible? In other words, can precarious researchers and underfunded institutions realistically afford such slow science approaches \citep{Bertone2026ai, Vostal2016accelerating}? 

The institutional decision to provide students and staff with access to these tools goes beyond economic concerns. AI-assisted coding, in particular, raises broader issues regarding infrastructural dependency and epistemic governance. The most popular LLMs are proprietary platforms with opaque model architectures that are certainly not designed or governed by the archaeological community. If we are working with sensitive or high-resolution datasets, such as genetic data, for example, reliance on external systems may create real issues in terms of institutional responsibility and data privacy. These issues are particularly pertinent when data are processed through commercial APIs that are not entirely transparent about their data storage or model training.

Another aspect of this discussion concerns the environmental costs of AI-assisted coding and AI use in general, as well as the infrastructural inequalities that could result from the proposed alternatives. To address data privacy and governance concerns, research institutions could adopt locally hosted or open-source models, such as those used for this research (e.g., \url{https://gwdg.de/en}). These models provide greater control over data flows and Retrieval-Augmented Generation (RAG) implementations. However, these solutions introduce a different form of inequality, as substantial computational resources are required to implement local LLM infrastructures, particularly on a departmental or institutional scale. Such systems may also be difficult to maintain and scale effectively. For example, if multiple researchers use them concurrently, shared local servers could become overloaded, raising broader questions about institutional investment, sustainability, and access. In practical terms, how much should universities be expected to invest in such infrastructures?

These questions are difficult to answer definitively because the solution depends on the purchasing power of each institution. Before committing to such investments, we as archaeological scientists might want to discuss a core issue: is the epistemic shift introduced by LLMs ultimately beneficial and should it be welcomed? The results presented in this paper suggest that underspecified prompts tend to produce suboptimal analytical choices. The risks go well beyond metacognitive laziness \citep{Fan2025beware} and sycophancy \citep{Li2026cognitive} and fundamentally create a further asymmetry between users who can audit a model’s output and those who cannot. With a vague request, an LLM does not encourage users to state their statistical assumptions, but it quietly converts hypothesis-driven analyses into open-ended data explorations, with analytical choices refined through interaction rather than reasoned in advance. 

While, in our view, LLMs can support hypothesis formation or critical reflection, their use has to be included in our scientific workflow rather than externalising this workflow as a single naive exploratory prompt. In particular, we define the scientific workflow as a complete causal inference process \citep{Bendixen2026cause, Gelman2026bayesian}, in which hypotheses, data collection, and modelling choices form part of a coherent inferential paradigm, rather than emerging through prompt-driven exploration.

In our perspective, LLMs are infrastructural tools that operate within the inferential process, but do not directly participate in it. Put simply, we think of LLMs in the same way as a calculator or a compiler. These tools do not alter statistical reasoning or programming logic, but still function to accelerate the execution of analytical choices. In the same way, the agentic use of AI tools should not externalise the epistemic responsibilities of the researcher.

\section{Conclusions and Prospects}\label{sec:conclusions}

The results of this study suggest that the introduction of ChatGPT in late 2022 and other large language models since then has not yet led to fewer diverse methods being used in archaeological literature. However, our concentration analysis indicates that the conditions for such a shift are already in place, particularly when users' engagement with these models is unclear, which could lead to measurable convergence in methods.

In this paper, we argue that the key issue surrounding the adoption of LLMs by archaeologists is not merely technological uptake, but rather the integration of these tools into existing scientific workflows. This is particularly salient with regard to the division of analytical responsibility between researchers and machines. The delegation of code writing to LLMs should fundamentally be understood not as a mere substitution of tools, but as a reconfiguration of epistemic practice, where the boundary between \emph{technē} (technical skill) and \emph{epistēmē} (theoretical reasoning) becomes increasingly blurred.

This reconfiguration has profound implications for transparency and reproducibility. AI-assisted analysis is particularly difficult to audit unless prompts, model versions and interaction histories are meticulously documented. However, even when prompts are disclosed, the stochastic nature of these tools prevents full reproducibility. In this context, emerging approaches such as `vibe engineering' or spec-driven development \citep{Piskala2026spec} may offer partial solutions by structuring LLM interactions in ways that improve transparency without limiting exploratory capacity. Rather than encouraging greater, uncritical reliance on LLMs, we aim to impose a formal structure on their use (see the difference between novice and expert prompts in \S\ref{sec:experiment_stream}, and in full in \S{}S6.4 of the Supplementary Materials). This type of formal structuring of prompts also has implications in educational contexts, where greater AI literacy has incentivised students to use LLMs more critically and produced higher levels of epistemic engagement \citep{Li2026cognitive}. Similarly, using open-source or locally hosted models can serve as a form of epistemic validation, providing greater control over data flows and analytical environments.

\section*{Declarations}

\textbf{Author Contributions.} Both authors contributed jointly to the conception of the study, the methods, the software and formal analysis, the visualisation, and the writing. The first author is based at the University of G\"ottingen; the second author, at the University of Cambridge, is the corresponding author. Both authors have reviewed and approved the submitted version.

\textbf{Funding.} This research received no specific grant from any funding agency in the public, commercial, or not-for-profit sectors.

\textbf{Competing Interests.} The authors declare no competing interests.

\section*{Data and Code Availability}\label{sec:data_code}

All analysis code is available at https://github.com/robertoragno/vibe-coding-archaeology, together with the Supplementary Materials cited throughout \S{}3. Every prompt used at every stage of the pipeline is reproduced verbatim in \S{}S4 and \S{}S6 of the Supplementary Materials, and every count reported in that document is re-derived from the released artefacts by an included verification script.

The raw Scopus metadata cannot be redistributed, as it is proprietary to Elsevier and covered by the terms of the API licence under which it was retrieved; the same applies to the source identifiers that define the corpus, which \S{}S3.4 instead shows how to regenerate from Elsevier's public Scopus Source List. Everything downstream is released: the retrieval script; the complete L1/L2/L3 taxonomy with member terms, frequencies and post-audit category assignments; the interactive taxonomy browser; the full stimulus set of the LLM experiment; and, for both models, every recommendation produced, with its justification, its consensus taxonomy mapping and all three raw mapping votes. A reader without Scopus access can therefore reproduce every analysis reported in \S\ref{sec:results} from the released artefacts alone.

\bibliographystyle{plainnat}
\bibliography{references}

\clearpage
\appendix
\addtocontents{toc}{\protect\setcounter{tocdepth}{2}}
\setcounter{secnumdepth}{0}
\setlength{\parskip}{0.6em}
\setlength{\parindent}{0pt}
\renewcommand{\headeright}{Supplementary Materials}
\renewcommand{\shorttitle}{Is AI reorienting archaeological methods?}

\begin{center}
  \vspace*{1.5em}
  {\LARGE \textbf{Supplementary Materials}}\\[1em]
  {\Large \textbf{(Whose defaults?) Is artificial intelligence reorienting archaeological methods?}}\\[1.5em]
  {\large Lorenzo Cardarelli\textsuperscript{1} \quad Roberto Ragno\textsuperscript{2,*}}\\[0.8em]
  {\small \textsuperscript{1}Seminar f\"ur Ur- und Fr\"uhgeschichte, Universit\"at G\"ottingen, Germany}\\[0.3em]
  {\small \textsuperscript{2}McDonald Institute for Archaeological Research, University of Cambridge, UK}\\[0.3em]
  {\small \textsuperscript{*}Corresponding author: \href{mailto:rr673@cam.ac.uk}{rr673@cam.ac.uk}}
\end{center}

\vspace{1.5em}
\tableofcontents
\newpage

This portion of ESM covers the machine-learning stream of the study: the retrieval of the Scopus corpus, the LLM-based extraction of computational methods from abstracts, the construction of the L1/\allowbreak{}L2/\allowbreak{}L3 method taxonomy, and the design and quality control of the LLM recommendation experiment.\allowbreak{} The Bayesian modelling stream (Dirichlet-Multinomial regression, concentration analysis, negative binomial regression, and the associated workflow and sensitivity checks) is documented in S7 to S12 below.\allowbreak{}

\section{S1: Computational environment}

Every LLM inference in this study was run locally, on consumer hardware, without any commercial API.\allowbreak{} This was a deliberate design choice: it makes the pipeline auditable and re-executable by any reader with comparable hardware, and it avoids the silent model-version drift that affects hosted endpoints, which would make a study of this kind impossible to reproduce even in principle.\allowbreak{}

\noindent\textbf{Table S1.1. Computational environment for the machine-learning stream.}

\begingroup\small
\begin{longtable}{>{\raggedright\arraybackslash}p{0.30\linewidth}>{\raggedright\arraybackslash}p{0.60\linewidth}}
\toprule
\textbf{Component} & \textbf{Specification} \\
\midrule\endfirsthead
\toprule
\textbf{Component} & \textbf{Specification} \\
\midrule\endhead
CPU & 13th Gen Intel Core i9-13900K \\
GPU & NVIDIA RTX A4000 (16 GB VRAM) \\
Operating system & Linux 7.\allowbreak{}0.\allowbreak{}0-28-generic, x86\_\allowbreak{}64, glibc 2.\allowbreak{}39 \\
Python & 3.\allowbreak{}12.\allowbreak{}13 (conda environment) \\
Inference backend & llama-cpp-python 0.\allowbreak{}3.\allowbreak{}18 (llama.\allowbreak{}cpp, GGUF) \\
Clustering & evoc 0.\allowbreak{}3.\allowbreak{}1, scikit-learn 1.\allowbreak{}8.\allowbreak{}0 \\
String matching & thefuzz 0.\allowbreak{}22.\allowbreak{}1 \\
Data handling & pandas 2.\allowbreak{}3.\allowbreak{}3, numpy 2.\allowbreak{}4.\allowbreak{}3, openpyxl 3.\allowbreak{}1.\allowbreak{}5 \\
Plotting & matplotlib 3.\allowbreak{}10.\allowbreak{}8 \\
\bottomrule
\end{longtable}
\endgroup

The Bayesian stream was fitted separately in R 4.\allowbreak{}6 via cmdstanr; see the docs/\allowbreak{} files listed above for its own environment statement.\allowbreak{}

\section{S2: Model cards and decoding settings}

Three models were used across the pipeline.\allowbreak{} The table below consolidates settings that are distributed across \S{}3.\allowbreak{}2 and \S{}3.\allowbreak{}4 of the main text, and adds the parameters not reported there.\allowbreak{}

\noindent\textbf{Table S2.1. Models and decoding settings, by pipeline stage.}

\begingroup\footnotesize
\begin{longtable}{>{\raggedright\arraybackslash}p{0.1\linewidth}>{\raggedright\arraybackslash}p{0.2\linewidth}>{\raggedright\arraybackslash}p{0.1\linewidth}>{\raggedright\arraybackslash}p{0.1\linewidth}>{\raggedright\arraybackslash}p{0.1\linewidth}>{\raggedright\arraybackslash}p{0.1\linewidth}>{\raggedright\arraybackslash}p{0.1\linewidth}>{\raggedright\arraybackslash}p{0.1\linewidth}}
\toprule
\textbf{Stage} & \textbf{Model} & \textbf{Quant.\allowbreak{}} & \textbf{n\_\allowbreak{}ctx} & \textbf{Temp.\allowbreak{}} & \textbf{Seed} & \textbf{Max tokens} & \textbf{Source} \\
\midrule\endfirsthead
\toprule
\textbf{Stage} & \textbf{Model} & \textbf{Quant.\allowbreak{}} & \textbf{n\_\allowbreak{}ctx} & \textbf{Temp.\allowbreak{}} & \textbf{Seed} & \textbf{Max tokens} & \textbf{Source} \\
\midrule\endhead
Method extraction (\S{}S4.\allowbreak{}1) & Qwen3.\allowbreak{}5-9B-Instruct & Q8\_\allowbreak{}0 & 4,096 & 0.\allowbreak{}01 & 42 & 150 & computational\_\allowbreak{}methods.\allowbreak{}py \\
Term embedding (\S{}S4.\allowbreak{}4) & Qwen3-Embedding-8B & Q8\_\allowbreak{}0 & 512 & --- & 42 & --- & build\_\allowbreak{}taxonomy\_\allowbreak{}supervised.\allowbreak{}py \\
L3 labelling (\S{}S4.\allowbreak{}5) & Qwen3.\allowbreak{}5-9B-Instruct & Q8\_\allowbreak{}0 & 4,096 & 0.\allowbreak{}1 & 42 & 150 & build\_\allowbreak{}taxonomy\_\allowbreak{}supervised.\allowbreak{}py \\
Garbage validation (\S{}S4.\allowbreak{}6) & Qwen3.\allowbreak{}5-9B-Instruct & Q8\_\allowbreak{}0 & 4,096 & 0.\allowbreak{}0 & 42 & 20 & build\_\allowbreak{}taxonomy\_\allowbreak{}supervised.\allowbreak{}py \\
\texorpdfstring{L3$\rightarrow$L2}{L3->L2} assignment (\S{}S4.\allowbreak{}7) & Qwen3.\allowbreak{}5-9B-Instruct & Q8\_\allowbreak{}0 & 4,096 & 0.\allowbreak{}0 & 42 & 20 & build\_\allowbreak{}taxonomy\_\allowbreak{}supervised.\allowbreak{}py \\
Adversarial review (\S{}S4.\allowbreak{}8) & Qwen3.\allowbreak{}5-9B-Instruct & Q8\_\allowbreak{}0 & 4,096 & 0.\allowbreak{}3 & 42 & 30 & build\_\allowbreak{}taxonomy\_\allowbreak{}supervised.\allowbreak{}py \\
Experiment --- generation (\S{}S6) & Qwen3.\allowbreak{}5-9B-Instruct /\allowbreak{} Gemma 4 E4B-it & Q8\_\allowbreak{}0 /\allowbreak{} UD-Q8\_\allowbreak{}K\_\allowbreak{}XL & 8,192 & 0.\allowbreak{}1 & --- & 1,024 & run\_\allowbreak{}experiment.\allowbreak{}py \\
Experiment --- \texorpdfstring{L4$\rightarrow$L3}{L4->L3} mapping (\S{}S6) & Qwen3.\allowbreak{}5-9B-Instruct /\allowbreak{} Gemma 4 E4B-it & Q8\_\allowbreak{}0 /\allowbreak{} UD-Q8\_\allowbreak{}K\_\allowbreak{}XL & 8,192 & 0.\allowbreak{}1 & --- & 128 & run\_\allowbreak{}experiment.\allowbreak{}py \\
\bottomrule
\end{longtable}
\endgroup

\section{S3: Scopus retrieval protocol}

This section discharges the commitments made in \S{}3.\allowbreak{}1 ("the custom Python script and the exact query parameters are provided in the supplementary materials") and \S{}3.\allowbreak{}2 ("for the complete list of extracted metadata, refer to the SM") of the main text.\allowbreak{}

\subsection{S3.1 Query construction}

Journals were identified from the Scopus Source List by the ASJC code 1204 (Archaeology), yielding 546 unique Scopus Source IDs.\allowbreak{} The identifiers themselves are Elsevier's and are not redistributed here; \S{}S3.\allowbreak{}4 gives the procedure for regenerating the identical list from the source, which takes a few minutes.\allowbreak{} Retrieval used the Scopus Search API endpoint https:/\allowbreak{}/\allowbreak{}api.\allowbreak{}elsevier.\allowbreak{}com/\allowbreak{}content/\allowbreak{}search/\allowbreak{}scopus with view=COMPLETE and a page size of 25.\allowbreak{}

For each source ID the script first issues a count query and then retrieves records with:

\begin{lstlisting}
SOURCE-ID(<source_id>)
\end{lstlisting}

When a source exceeds the API's result ceiling, the query is split by publication year:

\begin{lstlisting}
SOURCE-ID(<source_id>) AND PUBYEAR IS <year>
\end{lstlisting}

for each year from 2010 to 2025 (YEAR\_\allowbreak{}FROM = 2010, YEAR\_\allowbreak{}TO = datetime.\allowbreak{}now().\allowbreak{}year).\allowbreak{} Requests are throttled at 0.\allowbreak{}45 s and retried up to 5 times with a 5/\allowbreak{}10/\allowbreak{}20/\allowbreak{}40/\allowbreak{}60 s backoff; progress is checkpointed to progress.\allowbreak{}json per completed source, so retrieval is resumable.\allowbreak{} See Python/\allowbreak{}1\_\allowbreak{}dataset/\allowbreak{}downloader.\allowbreak{}py.\allowbreak{}

\subsection{S3.2 Metadata fields retrieved}

26 fields were requested from the API.\allowbreak{} The second column gives the corresponding column name in the exported scopus\_\allowbreak{}results.\allowbreak{}csv; a dash marks fields that were requested and are present in the raw API response but were not carried into the flat export (author affiliations, for instance, are retrieved as a nested structure and flattened into separate columns).\allowbreak{}

\noindent\textbf{Table S3.2. Scopus API metadata fields and their columns in the export.}

\begingroup\small
\begin{longtable}{>{\raggedright\arraybackslash}p{0.20\linewidth}>{\raggedright\arraybackslash}p{0.18\linewidth}>{\raggedright\arraybackslash}p{0.52\linewidth}}
\toprule
\textbf{\#} & \textbf{Scopus API field} & \textbf{Column in scopus\_\allowbreak{}results.\allowbreak{}csv} \\
\midrule\endfirsthead
\toprule
\textbf{\#} & \textbf{Scopus API field} & \textbf{Column in scopus\_\allowbreak{}results.\allowbreak{}csv} \\
\midrule\endhead
1 & dc:identifier & --- \\
2 & eid & eid \\
3 & dc:title & title \\
4 & dc:creator & first\_\allowbreak{}author \\
5 & author & --- \\
6 & prism:publicationName & source\_\allowbreak{}title \\
7 & prism:volume & volume \\
8 & prism:issueIdentifier & issue \\
9 & prism:pageRange & pages \\
10 & prism:coverDate & --- \\
11 & prism:doi & doi \\
12 & citedby-count & citation\_\allowbreak{}count \\
13 & subtypeDescription & doc\_\allowbreak{}type \\
14 & subtype & --- \\
15 & pubStatus & pub\_\allowbreak{}stage \\
16 & openaccess & open\_\allowbreak{}access \\
17 & affiliation & --- \\
18 & prism:issn & --- \\
19 & prism:isbn & isbn \\
20 & pubmed-id & pubmed\_\allowbreak{}id \\
21 & dc:publisher & publisher \\
22 & source-id & --- \\
23 & prism:aggregationType & source\_\allowbreak{}type \\
24 & authkeywords & --- \\
25 & dc:description & abstract \\
26 & language & --- \\
\bottomrule
\end{longtable}
\endgroup

\subsection{S3.3 From raw records to the analysis corpus}

\noindent\textbf{Table S3.3. From raw Scopus records to the analysis corpus.}

\begingroup\small
\begin{longtable}{>{\raggedright\arraybackslash}p{0.18\linewidth}>{\raggedright\arraybackslash}p{0.36\linewidth}>{\raggedright\arraybackslash}p{0.18\linewidth}>{\raggedright\arraybackslash}p{0.18\linewidth}}
\toprule
\textbf{Step} & \textbf{Operation} & \textbf{Source} & \textbf{Records} \\
\midrule\endfirsthead
\toprule
\textbf{Step} & \textbf{Operation} & \textbf{Source} & \textbf{Records} \\
\midrule\endhead
1 & Raw records retrieved from the Scopus Search API (2010--2025, 546 sources) & downloader.\allowbreak{}py & --- \\
2 & Deduplication on the Elsevier Identifier (eid) as primary key & downloader.\allowbreak{}py:280 & --- \\
3 & Restriction to doc\_\allowbreak{}type == "Article"; records with an empty authors or abstract field dropped & cleaning.\allowbreak{}py & 119,327 \\
4 & Articles with at least one computational method extracted & computational\_\allowbreak{}methods.\allowbreak{}py & 8,404 \\
5 & Articles retained after removal of the garbage cluster (\S{}S4.\allowbreak{}6) & build\_\allowbreak{}taxonomy\_\allowbreak{}supervised.\allowbreak{}py & 8,398 \\
\bottomrule
\end{longtable}
\endgroup

Steps 1 and 2 are left blank deliberately: the intermediate record counts depend on the state of the Scopus index at the time of retrieval and are not stable quantities.\allowbreak{} The reproducible figure is the cleaned corpus of 119,327 articles.\allowbreak{}

The 8,404 articles at step 4 contribute 17,229 (article, method) pairs, a mean of 2.\allowbreak{}05 extracted methods per article.\allowbreak{} This is consistent with the design rationale given in \S{}3.\allowbreak{}2 of the main text for preferring per-mention extraction over document-level topic modelling.\allowbreak{}

\subsection{S3.4 Regenerating the source list}

The 546 Scopus Source IDs are Elsevier identifiers and fall under the same licensing restriction as the metadata itself, so they are not redistributed with this repository.\allowbreak{} They are, however, fully regenerable from Elsevier's own publicly downloadable Scopus Source List, and the procedure is deterministic:

1.\allowbreak{}Download the Scopus Source List from https:/\allowbreak{}/\allowbreak{}www.\allowbreak{}scopus.\allowbreak{}com/\allowbreak{}sources.\allowbreak{}

2.\allowbreak{} Retain the rows whose All Science Journal Classification (ASJC) codes include 1204 (Archaeology).\allowbreak{}

3.\allowbreak{} Retain active journal titles and take the Sourcerecord ID column.\allowbreak{}

4.\allowbreak{} Write the identifiers one per line to Python/\allowbreak{}1\_\allowbreak{}dataset/\allowbreak{}source.\allowbreak{}txt, with the header line Sourcerecord ID, which is the input format downloader.\allowbreak{}py expects.\allowbreak{}

The corpus reported in this paper was retrieved on 29 April 2026, against the Source List current at that date.\allowbreak{} The Source List is revised periodically as titles are added, discontinued or reclassified, so a reader regenerating it later may obtain a slightly different set.\allowbreak{} This is a property of the source rather than of the procedure: the ASJC 1204 filter is the definition of the corpus, and 546 is the number of journals that satisfied it on the retrieval date.\allowbreak{}

\section{S4: Method extraction and taxonomy construction}

The prompts below are the epistemic core of the bibliometric stream: they are the operative definition of what this study counts as "a computational method" and of how methods are grouped.\allowbreak{} They are reproduced verbatim from the executed source, as Python source blocks, so that the f-string templates and the exact placeholder substitutions are both visible.\allowbreak{}

The main text (\S{}3.\allowbreak{}2) describes six phases.\allowbreak{} The pipeline as executed runs seven: a garbage validation pass (\S{}S4.\allowbreak{}6 below) sits between L3 labelling and \texorpdfstring{L3$\rightarrow$L2}{L3->L2} assignment.\allowbreak{} It is documented here because it is what reconciles the 242 clusters produced by the clustering step with the 241 clusters carried into the analysis.\allowbreak{}

\subsection{S4.1 Extraction of methods from abstracts}

One call per abstract, over the 119,327-article cleaned corpus.\allowbreak{} Defined at Python/\allowbreak{}2\_\allowbreak{}methods\_\allowbreak{}extractions/\allowbreak{}computational\_\allowbreak{}methods.\allowbreak{}py.\allowbreak{}

\begin{lstlisting}
def extract_computational_methods(abstract_text):
 	messages = [
     	{
         	"role": "system",
 "content": (
 	"You are a strict scientific classifier. Read the scientific abstract below "
 	"and identify ALL computational methods and techniques used in the research.\n\n"
 	"Rules:\n"
 	"1. List every specific computational or statistical technique, separated by ' | '.\n"
 	"2. Use the EXACT common name (e.g. 'Random Forest', 'PCA', 'k-means', 'LSTM', 'Kriging').\n"
 	"3. Match the level of specificity in the abstract: if it names a specific technique "
 	"(e.g. 'CNN', 'Random Forest'), use that name. If it only mentions a broad category "
 	"(e.g. 'deep learning', 'machine learning'), use that -- do NOT infer a specific method "
 	"that is not explicitly stated.\n"	
 	"4. Do NOT list software or languages (e.g. 'Python', 'R', 'QGIS', 'ArcGIS').\n"
 	"5. Use standard capitalization for well-known techniques.\n"
 	"6. If NO computational method is present, respond ONLY with: None\n"
 	"7. Output ONLY the technique name(s). No explanations, no categories.\n\n"
 	"Valid output examples:\n"
 	"  Random Forest | PCA\n"
 	"  Viewshed Analysis | KDE | k-means\n"
 	"  CNN | LSTM\n"
 	"  Deep Learning\n"
 	"  ANOVA\n"
 	"  None\n\n"
 )
 	
     	},
     	{
         	"role": "user",
         	"content": f"Abstract: {abstract_text}"
     	}
 	]
\end{lstlisting}

Post-processing of the response: surrounding quotes and stops are stripped, newlines collapsed; responses matching none, no, n/\allowbreak{}a, - or the empty string (case-insensitively) are normalised to the sentinel None; the remainder is split on | and each element is stripped.\allowbreak{} Abstracts shorter than 10 characters are assigned None without a model call.\allowbreak{}

\subsection{S4.2 Phase 1: Text normalisation}

Applied to the raw method strings before any similarity computation:

\begin{enumerate}
  \item British-to-American spelling conversion via a regular-expression dictionary (\textit{modelling} $\to$ \textit{modeling}, \textit{colour} $\to$ \textit{color}, \textit{visualisation} $\to$ \textit{visualization});
  \item removal of intra-word hyphens between alphabetic tokens;
  \item capitalisation standardisation, preserving all-uppercase acronyms and title-casing mixed-case terms;
  \item stripping of common English plurals from multi-character words.
\end{enumerate}

Rule 4 is aggressive and has a visible artefact in the released canonical vocabulary: terms ending in -sis are over-stripped, so Principal Component Analysis normalises to Principal Component Analysi.\allowbreak{} The deformation is applied uniformly, so it neither splits nor merges clusters that would otherwise differ, and the affected strings are still correctly grouped by the fuzzy and semantic stages that follow.\allowbreak{} It does mean that the canonical column of the released taxonomy should be read as an internal key, not as a display label; the human-readable names are the L3 labels of \S{}S5.\allowbreak{}2.\allowbreak{}

\subsection{S4.3 Phase 2: Fuzzy deduplication}

Normalised terms were grouped into equivalence classes with thefuzz's token\_\allowbreak{}sort\_\allowbreak{}ratio at a similarity threshold of 88/\allowbreak{}100.\allowbreak{} The metric is order-insensitive, so it collapses permutations such as Principal Component Analysis /\allowbreak{} Analysis Principal Component.\allowbreak{} Within each class the most frequent variant was retained as the canonical term.\allowbreak{}

\subsection{S4.4 Phase 3: Semantic embedding and L3 clustering}

\noindent\textbf{Table S4.4. Semantic embedding and L3 clustering settings.}

\begingroup\small
\begin{longtable}{>{\raggedright\arraybackslash}p{0.30\linewidth}>{\raggedright\arraybackslash}p{0.60\linewidth}}
\toprule
\textbf{Parameter} & \textbf{Value} \\
\midrule\endfirsthead
\toprule
\textbf{Parameter} & \textbf{Value} \\
\midrule\endhead
Embedding model & Qwen3-Embedding-8B, Q8\_\allowbreak{}0, n\_\allowbreak{}ctx=512 \\
Instruction prefix & ``Represent this computational method for grouping by its specific algorithmic paradigm and technique family, ignoring application domain'' \\
Clustering algorithm & EVoC (hierarchical, density-based) \\
Retained layer & layer 0 (finest-grained) \\
Minimum cluster size & 5 \\
Maximum hierarchy layers & 12 \\
Noise handling & points labelled $-$1 reassigned to the nearest cluster centroid (k-NN) \\
Final merge pass & pairs of clusters with centroid cosine similarity $\geq$ 0.\allowbreak{}98 fused iteratively \\
\bottomrule
\end{longtable}
\endgroup

The instruction prefix is doing substantive work: without it the embedding space organises terms by archaeological application domain (ceramics, landscape, bone) rather than by algorithmic family, which is the opposite of what the taxonomy needs.\allowbreak{}

\subsection{S4.5 Phase 4: Automatic L3 labelling}

Defined at Python/\allowbreak{}3\_\allowbreak{}classification/\allowbreak{}build\_\allowbreak{}taxonomy\_\allowbreak{}supervised.\allowbreak{}py.\allowbreak{}

\begin{lstlisting}
def generate_l3_label(member_terms: list) -> tuple[str, str]:
 	sample = ", ".join(member_terms[:30])
 	if len(member_terms) > 30:
     	sample += f" ... (+{len(member_terms) - 30} more)"
 	user_msg = (
     	f"These computational method terms form a fine-grained cluster:\n{sample}\n\n"
     	f"Provide:\n"
     	f"1. A SINGLE short label (2-5 words) naming the specific technique family.\n"
     	f"   GOOD: 'Convolutional Neural Networks', 'Bayesian Radiocarbon Calibration', "
     	f"'Elliptic Fourier Analysis'.\n"
     	f"   FORBIDDEN: generic terms like 'Methods', 'Techniques', 'Analysis', "
     	f"or compound labels joining two families with '&', '/' or 'and'.\n"
     	f"2. A concise description (1-2 sentences) of what specific techniques are grouped here.\n\n"
     	f"Reply ONLY with valid JSON: {{\"label\": \"...\", \"description\": \"...\"}}"
 	)
 	messages = [
     	{"role": "system", "content": (
         	"You are a taxonomy expert for computational methods. "
         	"Reply ONLY with valid JSON with fields 'label' and 'description'."
     	)},
     	{"role": "user", "content": user_msg},
 	]
\end{lstlisting}

\subsection{S4.6 Phase 4.5: Garbage validation (not described in the main text)}

Density-based clustering over a vocabulary this heterogeneous inevitably produces at least one residual cluster that absorbs terms with no methodological content.\allowbreak{} A single LLM pass, run over every L3 cluster with a strongly conservative prompt (garbage=false when in doubt), identifies them.\allowbreak{}

Defined at Python/\allowbreak{}3\_\allowbreak{}classification/\allowbreak{}build\_\allowbreak{}taxonomy\_\allowbreak{}supervised.\allowbreak{}py.\allowbreak{}

\begin{lstlisting}
def is_cluster_garbage(l3_label: str, member_terms: list) -> bool:
 	sample = ", ".join(member_terms[:40])
 	if len(member_terms) > 40:
     	sample += f" ... (+{len(member_terms) - 40} more)"
 	user_msg = (
     	f"L3 cluster:\n"
     	f"  Label: '{l3_label}'\n"
     	f"  Members: {sample}\n\n"
     	f"Does this cluster contain a SIGNIFICANT PROPORTION of terms that are NOT "
     	f"computational/statistical/analytical methods?\n\n"
     	f"Reply ONLY with valid JSON: {{\"garbage\": false}} or {{\"garbage\": true}}"
 	)
 	messages = [
     	{"role": "system", "content": (
         	"You are a garbage detector for taxonomy clusters of computational methods. "
         	"Mark garbage=TRUE only if a notable share of the members are NOT scientific "
         	"methods -- for example: generic words (Integration, Comparison, Control, Part, "
         	"Match, Zoom, Buffer), software product names, non-analytical activities "
         	"(Interview, Trial Excavation), or completely unrelated jargon mixed together "
         	"with no coherent methodological theme. "
         	"Mark garbage=FALSE if the members are mostly real scientific methods, even if "
         	"the cluster is broad, contains many terms, spans multiple sub-fields, or has "
         	"a generic-sounding label. "
         	"Examples of NOT garbage: 'Kernel Density Estimation', 'Support Vector Machine "
         	"Variants', 'Statistical Hypothesis Testing', 'Multivariate Statistical Methods', "
         	"'Audio Signal Processing', 'Redundancy Analysis'. "
         	"When in doubt, reply garbage=false. "
         	"Reply ONLY with valid JSON."
     	)},
     	{"role": "user", "content": user_msg},
 	]
\end{lstlisting}

Outcome.\allowbreak{} Of the 242 L3 clusters produced by the clustering step, exactly one was flagged: L3-241, auto-labelled "Archaeological Dating and Analysis Methods".\allowbreak{} Inspection confirms the flag: its members include Buffer, Comparison, Control, Choice, Coding, Cloning, Card Sorting and Companion, precisely the class of generic words the prompt enumerates.\allowbreak{} The cluster absorbs 119 canonical terms and touches 145 (article, method) pairs; removing it costs only 6 articles (8,404 $\rightarrow$ 8,398), because nearly every affected article also reports at least one genuine method.\allowbreak{}

The 241 L3 clusters reported throughout the main text are the 242 produced minus this one.\allowbreak{}

\subsection{S4.7 Phase 5: L3-to-L2 assignment}

Defined at Python/\allowbreak{}3\_\allowbreak{}classification/\allowbreak{}build\_\allowbreak{}taxonomy\_\allowbreak{}supervised.\allowbreak{}py.\allowbreak{}

\begin{lstlisting}
def assign_l3_to_l2(l3_label: str, l3_terms: list) -> int:

 	sample = ", ".join(l3_terms[:25])
 	if len(l3_terms) > 25:
     	sample += f" ... (+{len(l3_terms) - 25} more)"

 	user_msg = (
     	f"You are classifying computational methods used in archaeology into a fixed taxonomy.\n\n"
     	f"L3 cluster to classify:\n"
     	f"  Label: '{l3_label}'\n"
     	f"  Terms: {sample}\n\n"
     	f"Available L2 categories (choose exactly ONE):\n{L2_CATALOG}\n\n"
     	f"Instructions:\n"
     	f"  - Choose the category whose description BEST matches the methodological paradigm "
     	f"of the L3 cluster, not the archaeological application.\n"
     	f"  - Focus on the algorithmic/statistical nature of the methods, not on what they "
     	f"are applied to (e.g. ceramics, sites, texts).\n"
     	f"  - If uncertain between two categories, prefer the more specific one.\n\n"
     	f"Reply ONLY with valid JSON: {{\"l2_index\": <number in brackets above>}}"
 	)
 	messages = [
     	{"role": "system", "content": (
         	"You are a taxonomy expert for computational methods in archaeology. "
         	"Reply ONLY with valid JSON: {\"l2_index\": N} where N is the number "
         	"in brackets from the category list."
     	)},
     	{"role": "user", "content": user_msg},
 	]
\end{lstlisting}

L2\_\allowbreak{}CATALOG is the full catalogue of the 25 L2 categories, each rendered as [index] Name: description, using the definitions listed in \S{}S5.\allowbreak{}1.\allowbreak{} Recovery cascade: strict JSON parse of \{"l2\_\allowbreak{}index": N\}; failing that, the first integer in the response that falls within the valid index range; failing that, -1 (unassigned).\allowbreak{} No cluster required the -1 fallback: all 242 received a valid L2 index.\allowbreak{}

\subsection{S4.8 Phase 6: Adversarial review}

Defined at Python/\allowbreak{}3\_\allowbreak{}classification/\allowbreak{}build\_\allowbreak{}taxonomy\_\allowbreak{}supervised.\allowbreak{}py.\allowbreak{}

\begin{lstlisting}
def review_l3_assignment(l3_id: int, l3_label: str, l3_terms: list,
                           current_l2_idx: int, n_votes: int = 3) -> int:

 	if current_l2_idx == -1:
     	return -1
 	current_l2_name = L2_NAMES_LIST[current_l2_idx]
 	current_l2_desc = L2_TAXONOMY[current_l2_name]
 	sample = ", ".join(l3_terms[:20])

 	user_msg = (
     	f"Audit task: verify the placement of a computational methods cluster.\n\n"
     	f"L3 cluster:\n"
     	f"  Label: '{l3_label}'\n"
     	f"  Terms: {sample}\n\n"
     	f"Currently assigned to L2 [{current_l2_idx}] '{current_l2_name}':\n"
     	f"  {current_l2_desc}\n\n"
     	f"Is this assignment correct?\n"
     	f"- If YES -> reply {{\"correct\": true}}\n"
     	f"- If NO, and the current category is COMPLETELY UNRELATED to the cluster's "
     	f"methodology -> reply {{\"correct\": false, \"better_l2_index\": <N>}}\n\n"
     	f"All L2 categories for reference:\n{L2_CATALOG}\n\n"
     	f"Reply ONLY with valid JSON."
 	)
 	messages = [
     	{"role": "system", "content": (
         	"You are a conservative taxonomy auditor for computational methods in archaeology. "
         	"Your default answer is {\"correct\": true}. "
         	"Only flag a misplacement when the current category is COMPLETELY UNRELATED "
         	"to the cluster's core methodology -- not because another category seems "
         	"marginally or equally good. If the current assignment is plausible or "
         	"defensible from any reasonable angle, reply correct=true. "
         	"When in doubt, reply correct=true. "
         	"Reply ONLY with valid JSON."
     	)},
     	{"role": "user", "content": user_msg},
 	]
\end{lstlisting}

The design is deliberately asymmetric: a reassignment is applied only when all three votes agree unanimously on the same alternative category, so a single dissenting vote preserves the original assignment.\allowbreak{} As noted in \S{}S2, the temperature of 0.\allowbreak{}3 is what makes the three votes independent enough for the unanimity rule to carry information.\allowbreak{}

\section{S5: The full taxonomy}

The hierarchy is asymmetric by design.\allowbreak{} L1 (7 thematic domains) is an editorial reading aid with no computational role: it exists only to group the L2 table below into recognisable sub-disciplines.\allowbreak{} L2 (25 categories) is researcher-defined and fixed in advance, with each category given a definition and representative methods that are injected verbatim into the assignment and review prompts of \S{}S4.\allowbreak{}7 through \S{}S4.\allowbreak{}8.\allowbreak{} L3 (241 clusters) is fully data-driven: discovered from the corpus, never enumerated in advance, and labelled automatically.\allowbreak{}

\subsection{S5.1 The 25 L2 categories}

Counts are over the analysis corpus, excluding the garbage cluster.\allowbreak{} Mentions counts (article, method) pairs; Articles counts distinct articles, so an article reporting two methods from the same L2 is counted once.\allowbreak{}

\noindent\textbf{Table S5.1. The 25 L2 categories, with definitions and corpus counts.}

\begingroup\footnotesize
\begin{longtable}{>{\raggedright\arraybackslash}p{0.113\linewidth}>{\raggedright\arraybackslash}p{0.113\linewidth}>{\raggedright\arraybackslash}p{0.225\linewidth}>{\raggedright\arraybackslash}p{0.113\linewidth}>{\raggedright\arraybackslash}p{0.113\linewidth}>{\raggedright\arraybackslash}p{0.113\linewidth}>{\raggedright\arraybackslash}p{0.113\linewidth}}
\toprule
\textbf{L1 domain} & \textbf{L2 category} & \textbf{Definition} & \textbf{L3 clusters} & \textbf{Canonical terms} & \textbf{Mentions} & \textbf{Articles} \\
\midrule\endfirsthead
\toprule
\textbf{L1 domain} & \textbf{L2 category} & \textbf{Definition} & \textbf{L3 clusters} & \textbf{Canonical terms} & \textbf{Mentions} & \textbf{Articles} \\
\midrule\endhead
Statistics \& Modeling & Univariate \& Classical Hypothesis Testing & Classical frequentist tests for comparing groups or distributions on a single variable, forming the baseline of inferential statistics before multivariate and computational methods.\allowbreak{} Examples: t-test, Mann-Whitney U, Chi-square, ANOVA, Kruskal-Wallis, Kolmogorov-Smirnov, Fisher's Exact Test, Wilcoxon Signed-Rank.\allowbreak{} & 15 & 397 & 1299 & 1015 \\
Statistics \& Modeling & Multivariate Analysis \& Dimensionality Reduction & Techniques that reduce or decompose high-dimensional datasets into lower-dimensional representations, including ordination, matrix factorization and manifold learning.\allowbreak{} Examples: PCA, Correspondence Analysis, MDS, t-SNE, Procrustes Analysis, Nonlinear Manifold Learning, Factor Analysis.\allowbreak{} & 11 & 407 & 1775 & 1600 \\
Statistics \& Modeling & Regression \& Generalized Linear Models & Parametric models that quantify relationships between a response variable and one or more predictors, including linear, logistic, Poisson and mixture variants.\allowbreak{} Examples: OLS Regression, GLM, GAM, Logistic Regression, Mixture Modeling, Generalized Additive Models.\allowbreak{} & 18 & 321 & 1013 & 873 \\
Statistics \& Modeling & Clustering \& Unsupervised Learning & Algorithms that partition unlabelled data into groups based on similarity, without prior class definitions.\allowbreak{} Examples: k-means, Hierarchical Clustering, DBSCAN, HDBSCAN, Gaussian Mixture Models, Self-Organizing Maps.\allowbreak{} & 6 & 111 & 529 & 501 \\
Statistics \& Modeling & Distance \& Similarity Metrics & Mathematical measures of dissimilarity or proximity between observations or features, used as building blocks for classification, clustering and retrieval.\allowbreak{} Examples: Mahalanobis Distance, Nearest Neighbor Analysis, Relief Algorithms, Cluster Validation Indices, Jaccard Similarity, Edit Distance.\allowbreak{} & 7 & 154 & 338 & 281 \\
\bottomrule
\end{longtable}
\endgroup

\subsection{S5.2 The 241 L3 clusters}

Ordered by mentions, descending.\allowbreak{} Terms is the number of distinct canonical terms absorbed by the cluster; the last column lists the five most frequent, as an inspection aid for judging whether the cluster coheres.\allowbreak{} On the appearance of strings such as Principal Component Analysi, see \S{}S4.\allowbreak{}2.\allowbreak{}

\noindent\textbf{Table S5.2. The 241 L3 clusters, ordered by mentions.}

\begingroup\footnotesize
\begin{longtable}{>{\raggedright\arraybackslash}p{0.129\linewidth}>{\raggedright\arraybackslash}p{0.129\linewidth}>{\raggedright\arraybackslash}p{0.129\linewidth}>{\raggedright\arraybackslash}p{0.129\linewidth}>{\raggedright\arraybackslash}p{0.129\linewidth}>{\raggedright\arraybackslash}p{0.257\linewidth}}
\toprule
\textbf{L3} & \textbf{Label} & \textbf{L2} & \textbf{Terms} & \textbf{Mentions} & \textbf{Most frequent member terms} \\
\midrule\endfirsthead
\toprule
\textbf{L3} & \textbf{Label} & \textbf{L2} & \textbf{Terms} & \textbf{Mentions} & \textbf{Most frequent member terms} \\
\midrule\endhead
L3-019 & Principal Component Analysis Variants & L2-01 & 36 & 571 & Principal Component Analysi, PCA, Principal Component, Minimum Noise Fraction, Principal Coordinate Analysi \\
L3-158 & Bayesian Statistical Inference & L2-05 & 69 & 557 & Bayesian Modeling, Bayesian Analysi, Bayesian Statistic, Bayesian Statistical Framework, Bayesian Approach \\
L3-226 & Geographic Information System Modeling & L2-09 & 30 & 448 & GIS, Geographic Information System, GIS Mapping, GIS Modeling, 3D GIS \\
L3-052 & K-Means Clustering & L2-03 & 49 & 352 & Cluster Analysi, Clustering, k-means, k-means Clustering, Clustering Algorithm \\
L3-232 & Specialized Domain Analyses & L2-01 & 151 & 343 & Content Analysi, Redundancy Analysi, Meta Analysi, Mediation Analysi, SWOT Analysi \\
L3-183 & Geometric Morphometrics & L2-13 & 62 & 333 & Geometric Morphometric, Geometric Morphometric Analysi, Morphometric Analysi, Traditional Morphometric, 3D Geometric Morphometric Method \\
L3-236 & Automated Content Classification & L2-15 & 73 & 256 & Machine Learning, Supervised Classification, Classification, Data Mining, Artificial Intelligence \\
L3-084 & Scanning Electron Microscopy with Energy Dispersive Spectroscopy & L2-18 & 34 & 249 & SEM EDS, SEM EDX, Scanning Electron Microscopy, Scanning Electron Microscopy With Energy Dispersive X-ray Spectroscopy, EDX Microanalysi \\
\bottomrule
\end{longtable}
\endgroup

\section{S6: LLM experiment: design and quality control}

The design rationale is given in \S{}3.\allowbreak{}4 of the main text and at greater length in experiment/\allowbreak{}README.\allowbreak{}md.\allowbreak{} This section supplies the full stimulus set, the prompts as executed, and the quality-control figures.\allowbreak{}

\subsection{S6.1 The 28 research questions}

One per thematic category of the Scopus corpus, formulated in substantive terms without naming or implying any computational solution.\allowbreak{} Lexical neutrality matters most for the novice profile, where the question is the model's only input and any method word in the stem would prime the recommendation directly.\allowbreak{}

\noindent\textbf{Table S6.1. The 28 research questions, one per thematic category.}

\begingroup\small
\begin{longtable}{>{\raggedright\arraybackslash}p{0.20\linewidth}>{\raggedright\arraybackslash}p{0.18\linewidth}>{\raggedright\arraybackslash}p{0.52\linewidth}}
\toprule
\textbf{\#} & \textbf{Thematic category} & \textbf{Research question} \\
\midrule\endfirsthead
\toprule
\textbf{\#} & \textbf{Thematic category} & \textbf{Research question} \\
\midrule\endhead
1 & Artefacts /\allowbreak{} Finds & How do I classify and interpret a heterogeneous assemblage of artefacts recovered from an archaeological context? \\
2 & Excavation /\allowbreak{} Survey report & How do I systematically document and communicate the results of an excavation or surface survey? \\
3 & Architecture and other features & How do I analyse and interpret architectural structures or built features at a site? \\
4 & Burials /\allowbreak{} Human remains & How do I study funerary practices and biological characteristics of a past population from skeletal remains? \\
5 & Period /\allowbreak{} Tradition discussion & How do I characterise and compare material cultures from different periods or traditions to identify continuities and discontinuities? \\
6 & Approaches /\allowbreak{} Theories /\allowbreak{} Methodology & How do I evaluate the effectiveness of a theoretical or methodological approach applied to a specific archaeological problem? \\
7 & Site(s) discussion & How do I interpret the function, chronology, and overall significance of an archaeological site? \\
8 & Art history /\allowbreak{} Iconography & How do I systematically analyse and interpret images, symbols, or figurative representations in an archaeological context? \\
9 & Tablet find /\allowbreak{} Texts /\allowbreak{} Inscriptions /\allowbreak{} Philology & How do I extract historical and cultural information from a corpus of ancient epigraphic or textual sources? \\
10 & Palaeoenvironment /\allowbreak{} Geoarchaeology /\allowbreak{} Geology & How do I reconstruct the environmental and geomorphological conditions in which past human activities took place? \\
11 & Political /\allowbreak{} Economic /\allowbreak{} Social Organisation & How do I reconstruct the social, economic, or political structure of a community from material evidence? \\
12 & Zooarchaeology & How do I analyse faunal remains to reconstruct hunting, herding practices, and human-animal relationships? \\
13 & Resource exploitation /\allowbreak{} Manufacture /\allowbreak{} Technology & How do I reconstruct the chaîne opératoire and raw material processing techniques of past societies? \\
14 & Ritual /\allowbreak{} Cult /\allowbreak{} Myths /\allowbreak{} Religion & How do I identify and interpret ritual or religious behaviour from material and contextual evidence? \\
15 & Evental History /\allowbreak{} Historical Geography & How do I integrate written historical sources and material data to reconstruct past events and geographical transformations? \\
16 & Subsistence economy /\allowbreak{} Food /\allowbreak{} Diet & How do I reconstruct the dietary and subsistence strategies of a past community? \\
17 & Urban archaeology /\allowbreak{} Urbanism & How do I analyse the organisation, growth, and transformation of an ancient urban context? \\
18 & Archaeometry & How do I determine the composition, provenance, or manufacturing techniques of an artefact through physicochemical analysis? \\
19 & Chronology /\allowbreak{} Dating & How do I build a reliable chronological sequence when available dating evidence is uncertain or fragmentary? \\
20 & History of archaeological research & How do I reconstruct the evolution of approaches and research interests within a discipline over time? \\
21 & Landscape /\allowbreak{} Settlement /\allowbreak{} Territorial studies & How do I analyse the distribution and organisation of human settlements across a territory over the long term? \\
22 & Archaeobotany /\allowbreak{} Palynology & How do I reconstruct past vegetation, plant use, and landscape change through botanical remains? \\
23 & Textile /\allowbreak{} Textile tools & How do I analyse textile production and its economic and cultural significance in a past society? \\
24 & Architectural decorations & How do I document, classify, and interpret architectural decorative programmes within a historical and cultural context? \\
25 & Trade /\allowbreak{} Exchange /\allowbreak{} Interactions & How do I reconstruct exchange networks and cultural interaction between distant communities? \\
26 & Heritage /\allowbreak{} Conservation & How do I assess, manage, and communicate the value of a cultural asset within a protection and risk framework? \\
27 & Experimental archaeology /\allowbreak{} Ethnoarchaeology & How do I use comparisons with modern or experimental practices to interpret material evidence from the past? \\
28 & Rock art & How do I document, classify, and interpret rock art manifestations within their spatial and cultural context? \\
\bottomrule
\end{longtable}
\endgroup

\subsection{S6.2 The 9 vague methodological descriptors}

These stand for the informal prior knowledge of a researcher with partial computational training.\allowbreak{} They are deliberately imprecise and do not map one-to-one onto any L2 category, so the intermediate profile must bridge from an informal description to a concrete technique.\allowbreak{}

\noindent\textbf{Table S6.2. The nine vague methodological descriptors.}

\begingroup\small
\begin{longtable}{>{\raggedright\arraybackslash}p{0.30\linewidth}>{\raggedright\arraybackslash}p{0.60\linewidth}}
\toprule
\textbf{\#} & \textbf{Vague methodological descriptor} \\
\midrule\endfirsthead
\toprule
\textbf{\#} & \textbf{Vague methodological descriptor} \\
\midrule\endhead
1 & some kind of statistical or quantitative analysis \\
2 & some kind of spatial or geographic approach \\
3 & some kind of image or visual analysis \\
4 & some kind of network or relational analysis \\
5 & some kind of text or document analysis \\
6 & some kind of dating or chronological modelling \\
7 & some kind of 3D reconstruction or modelling \\
8 & some kind of machine learning or pattern recognition \\
9 & some kind of simulation or agent-based modelling \\
\bottomrule
\end{longtable}
\endgroup

The 25 L2 category names of \S{}S5.\allowbreak{}1 serve as the expert-profile input, supplied as the bare category name.\allowbreak{}

\subsection{S6.3 Sampling design}

Each of the 252 iterations (28 questions $\times$ 9 vague descriptors) samples one triplet: one research question, one L2 category, and one vague descriptor.\allowbreak{} It submits this triplet to all three profiles.\allowbreak{} The three profiles within an iteration therefore face identical archaeological content and differ only in methodological framing, which is what licenses the within-iteration comparison.\allowbreak{} Total generation calls: 252 $\times$ 3 = 756 per model, plus three mapping calls per extracted L4 method.\allowbreak{}

\subsection{S6.4 The three profile prompts}

Defined at Python/\allowbreak{}4\_\allowbreak{}experiment/\allowbreak{}run\_\allowbreak{}experiment.\allowbreak{}py, the system framing and the output-format instruction are byte-identical across the three; only the researcher's stated prior knowledge varies.\allowbreak{}

\begin{lstlisting}
def prompt_expert(question: str, l2: str) -> str:
 	return (
     	"You are a research assistant for computational archaeologists.\n\n"
     	"A researcher comes to you with the following problem:\n\n"
     	f'"I am working on: {question}\n'
     	f'I already know I want to apply {l2} to my analysis.\n'
     	"Which specific tools, algorithms, or variants of this method would you\n"
     	'recommend, and how would you apply them concretely to this problem?"\n\n'
     	"List the computational methods you would use, being as specific as possible.\n"
     	"For each method, provide a one-sentence justification.\n"
     	"Format each line strictly as:  Method Name | One-sentence justification."
 	)


 def prompt_intermediate(question: str, vague: str) -> str:
 	return (
     	"You are a research assistant for computational archaeologists.\n\n"
     	"A researcher comes to you with the following problem:\n\n"
     	f'"I am working on: {question}\n'
     	f'I have a rough idea that I need {vague},\n'
     	'but I do not know which specific method to choose.\n'
     	'What would you recommend?"\n\n'
     	"List the computational methods you would use, being as specific as possible.\n"
     	"For each method, provide a one-sentence justification.\n"
     	"Format each line strictly as:  Method Name | One-sentence justification."
 	)


 def prompt_novice(question: str) -> str:
 	return (
     	"You are a research assistant for computational archaeologists.\n\n"
     	"A researcher comes to you with the following problem:\n\n"
     	f'"I am working on: {question}\n'
     	"I have no specific computational background.\n"
     	'Which digital methods could I use to address this research problem?"\n\n'
     	"List the computational methods you would use, being as specific as possible.\n"
     	"For each method, provide a one-sentence justification.\n"
     	"Format each line strictly as:  Method Name | One-sentence justification."
 	)
\end{lstlisting}

\subsection{S6.5 The \texorpdfstring{L4$\rightarrow$L3}{L4->L3} mapping prompt}

Defined at Python/\allowbreak{}4\_\allowbreak{}experiment/\allowbreak{}run\_\allowbreak{}experiment.\allowbreak{}py.\allowbreak{} Run three independent times per extracted L4 item.\allowbreak{}

\begin{lstlisting}
def prompt_l4_to_l3(l4_method: str, l3_list: list) -> str:
 	taxonomy_block = "\n".join(l3_list)
 	return (
     	"You are a taxonomy classifier.\n\n"
     	"Given the following L4 computational method, identify the single most appropriate\n"
     	"L3 category from the list below.  Reply with ONLY the exact L3 label as it appears\n"
     	"in the list -- nothing else, no explanation.\n\n"
     	f"L4 method: {l4_method}\n\n"
     	"L3 taxonomy:\n"
     	f"{taxonomy_block}\n\n"
     	"Best matching L3 category:"
 	)
\end{lstlisting}

Label extraction from the response follows a priority cascade: exact string match against the L3 list; numeric code prefix match (e.\allowbreak{}g.\allowbreak{} L3-097); longest substring match against the descriptive part of any label; and, as a last resort, the raw response truncated to 300 characters, automatically flagged as inconsistent.\allowbreak{} The consensus label is the majority vote over the three runs; the item is flagged consistent only when all three agree.\allowbreak{}

Each generation response is parsed line by line into (method, justification) pairs, tolerating numbered lists, bullets and bold markers.\allowbreak{} Responses yielding no parseable item are stored as a single UNPARSED row rather than silently dropped.\allowbreak{}

\subsection{S6.6 Quality control}

\noindent\textbf{Table S6.6a. \texorpdfstring{L4$\rightarrow$L3}{L4->L3} mapping consistency, by model.}

\begingroup\footnotesize
\begin{longtable}{>{\raggedright\arraybackslash}p{0.129\linewidth}>{\raggedright\arraybackslash}p{0.129\linewidth}>{\raggedright\arraybackslash}p{0.257\linewidth}>{\raggedright\arraybackslash}p{0.129\linewidth}>{\raggedright\arraybackslash}p{0.129\linewidth}>{\raggedright\arraybackslash}p{0.129\linewidth}}
\toprule
\textbf{Model} & \textbf{Rows (L4 items)} & \textbf{Consistent mappings} & \textbf{Agreement rate} & \textbf{Flagged for review} & \textbf{UNPARSED} \\
\midrule\endfirsthead
\toprule
\textbf{Model} & \textbf{Rows (L4 items)} & \textbf{Consistent mappings} & \textbf{Agreement rate} & \textbf{Flagged for review} & \textbf{UNPARSED} \\
\midrule\endhead
Qwen & 6,915 & 6,651 & 96.\allowbreak{}2\% & 264 & 0 \\
Gemma & 5,441 & 5,252 & 96.\allowbreak{}5\% & 189 & 0 \\
\bottomrule
\end{longtable}
\endgroup

\noindent\textbf{Table S6.6b. L4 method items by researcher profile.}

\begingroup\footnotesize
\begin{longtable}{>{\raggedright\arraybackslash}p{0.15\linewidth}>{\raggedright\arraybackslash}p{0.15\linewidth}>{\raggedright\arraybackslash}p{0.3\linewidth}>{\raggedright\arraybackslash}p{0.15\linewidth}>{\raggedright\arraybackslash}p{0.15\linewidth}}
\toprule
\textbf{Model} & \textbf{Novice} & \textbf{Intermediate} & \textbf{Expert} & \textbf{Total} \\
\midrule\endfirsthead
\toprule
\textbf{Model} & \textbf{Novice} & \textbf{Intermediate} & \textbf{Expert} & \textbf{Total} \\
\midrule\endhead
Qwen & 2,491 & 2,222 & 2,202 & 6,915 \\
Gemma & 1,800 & 1,724 & 1,917 & 5,441 \\
\bottomrule
\end{longtable}
\endgroup

\subsection{S6.7 Coverage of the taxonomy by the recommendations}

The raw method strings produced by the two models collapse almost entirely onto the existing L3 vocabulary.\allowbreak{} Qwen produced 2,904 distinct L4 strings (after normalising case and whitespace), Gemma 1,746; these map onto 194 and 167 distinct L3 labels respectively.\allowbreak{}

Two clarifications on those figures, which are reported in \S{}3.\allowbreak{}4 of the main text:

The denominator is the 242-label taxonomy, not the 241 analysable labels.\allowbreak{} Both models mapped some items into the garbage cluster L3-241, so against the 241 analysable labels the coverage figures are 193 and 166.\allowbreak{}

Unlike the analyses in \S{}3.\allowbreak{}4.\allowbreak{}1 through \S{}3.\allowbreak{}4.\allowbreak{}2, this check uses all mapping attempts rather than the majority-consistent subset, since detecting out-of-taxonomy output does not require cross-run agreement.\allowbreak{}

Exactly one mapping attempt fell outside the taxonomy, in the Gemma run: the string L3-27, which is a malformed rendering of an L3 code (the taxonomy uses three-digit codes, L3-027).\allowbreak{} Footnote 2 of the main text attributes this to a malformed L2 label; it is an L3 label.\allowbreak{} Source: data/\allowbreak{}output/\allowbreak{}recirculation\_\allowbreak{}stats.\allowbreak{}csv, produced by R/\allowbreak{}experiment/\allowbreak{}06\_\allowbreak{}concentration.\allowbreak{}R.\allowbreak{}

\section{S7: The Dirichlet-Multinomial model}

This section documents the model described in \S{}3.\allowbreak{}3 of the main text: priors, sampler settings, convergence diagnostics, and two implementation details not shown in the model equation.

\subsection{S7.1 Model specification}
We model method counts compositionally within each L2 sub-discipline. For sub-discipline \(g\), year \(t\), and technique \(k\), where \(K_g\) is the number of techniques in sub-discipline \(g\), the model is

\begin{align}
\eta_{g,t,k} &=
\mu_{g,k}
+ \beta_{g,k}\,\mathrm{year}_{\mathrm{std},t}
+ \gamma_{g,k}\,\mathrm{postLLM}_{t}, \\
\boldsymbol{\alpha}_{g,t} &=
\phi\,\operatorname{softmax}\!\left(\boldsymbol{\eta}_{g,t}\right), \\
\mathbf{y}_{g,t} &\sim
\operatorname{DirichletMultinomial}
\!\left(\boldsymbol{\alpha}_{g,t}\right).
\end{align}

where \(\mu_{g,k}\) is the baseline log-share of technique \(k\), \(\beta_{g,k}\) its linear trend per standardised year, \(\gamma_{g,k}\) the post-LLM level shift from 2023, and \(\phi\) the Dirichlet-Multinomial concentration (higher = proportions tighter around the expected composition).

The $\gamma_{g,k}$ parameters therefore represent a change in level relative to the pre-existing linear trend defined by $\beta_{g,k}$. With only three years of post-2023 observations, the model cannot distinguish an abrupt change from a gradual acceleration, and we do not attempt such an interpretation in the main text. The primary estimand is $\sigma_{\gamma}$, which represents the across-technique scale of these deviations, rather than any individual $\gamma_{g,k}$.

The Dirichlet--Multinomial likelihood is implemented directly as the \texttt{dm\_log()} function in the \texttt{functions} block of \texttt{stan/bibliometric\_dirichlet\_multinomial.stan}. For a vector of counts $\mathbf{n}$ and Dirichlet parameters $\boldsymbol{\alpha}$, the log-likelihood is:

\begin{equation}
\begin{aligned}
\log p(\mathbf{n}\mid\boldsymbol{\alpha})
={}&
\log\Gamma\left(\alpha_0\right)
-\log\Gamma\left(\alpha_0+N\right) \\
&+
\sum_{k=1}^{K}
\left[
\log\Gamma\left(\alpha_k+n_k\right)
-\log\Gamma\left(\alpha_k\right)
\right],
\end{aligned}
\end{equation}

where

\begin{equation}
\alpha_0 = \sum_{k=1}^{K}\alpha_k,
\qquad
N = \sum_{k=1}^{K}n_k.
\end{equation}

We omit group--year cells with no observations rather than contributing a zero-count likelihood term.

\subsection{S7.2 Priors and implementation details}

\noindent\textbf{Table S7.1. Priors on the Dirichlet-Multinomial model.}

\begingroup\small
\begin{longtable}{>{\raggedright\arraybackslash}p{0.20\linewidth}>{\raggedright\arraybackslash}p{0.18\linewidth}>{\raggedright\arraybackslash}p{0.52\linewidth}}
\toprule
\textbf{Parameter} & \textbf{Prior} & \textbf{Rationale} \\
\midrule
\endfirsthead

\toprule
\textbf{Parameter} & \textbf{Prior} & \textbf{Rationale} \\
\midrule
\endhead

$\log \phi$
&
$\operatorname{Normal}(\log(100), 1)$
&
Equivalent to $\phi \sim \operatorname{Lognormal}(\log(100), 1)$, giving a median of 100, a mean of approximately 165, and an approximate 95\% interval of $[14, 710]$. Modelling $\log \phi$ avoids the lower bound at zero.
\\

$\sigma_{\beta}$
&
$\operatorname{Exponential}(2)$
&
Controls the variation in baseline time trends across techniques over the 16-year period.
\\

$\sigma_{\gamma}$
&
$\operatorname{Exponential}(4)$
&
A more restrictive prior than for $\sigma_{\beta}$, reflecting the shorter post-2023 period and the expectation that these deviations are smaller than the baseline trends estimated over the full study period.
\\

$\mu_{\mathrm{raw}},\ \beta_{\mathrm{raw}},\ \gamma_{\mathrm{raw}}$
&
$\operatorname{Normal}(0, 1)$
&
We use standard normal priors for the raw parameters, with scale controlled by the corresponding $\sigma$, giving a non-centred parameterisation.
\\

\bottomrule
\end{longtable}
\endgroup

Earlier specifications fixed $\phi$ at 10 and then at 50; the posterior under the current model concentrates near 1,100, far above either value, so those fixed-$\phi$ runs were attributing structural variation to noise and are superseded.\allowbreak{}

The asymmetry between the two hyperpriors is intentional: it makes a large $\sigma_\gamma$ harder to obtain, so the reported posterior is not an artefact of a permissive prior.\allowbreak{} The cost is that individual $\gamma$ estimates carry substantial uncertainty, which is why we report inference at the level of the scale parameter and of directional patterns rather than for single techniques.\allowbreak{}

Two additional features of the model are not apparent from the equation above but are important for its implementation.

\textbf{Soft sum-to-zero constraint.} Within each sub-discipline, the log-shares are identified only up to an additive constant. Rather than choosing one technique as a reference category, we use a common \texttt{Stan} feature, a weak sum-to-zero constraint\footnote{\url{https://mc-stan.org/docs/stan-users-guide/regression.html\#parameterizing-centered-vectors}} on $\mu$, $\beta$, and $\gamma$:

\begin{equation}
\sum_{k=1}^{K_g} \mu_{\mathrm{raw},g,k} \sim \mathcal{N}(0,\ 0.001 \times K_g)
\end{equation}

This avoids assigning a special role to any one technique while allowing the constraint to remain weak rather than fixing the sum exactly.

\textbf{Pinned padding slots.} The count array is rectangular, with dimensions $N_{\mathrm{groups}} \times N_{\mathrm{years}} \times K_{\max}$, whereas the number of techniques varies between sub-disciplines. We set $K_{\max}=20$ and use the remaining slots only as padding. These unused parameters are given tight priors around zero:

\begin{equation}
  \mu_{\mathrm{raw},g,k},\ \beta_{\mathrm{raw},g,k},\ \gamma_{\mathrm{raw},g,k}
  \sim \mathcal{N}(0,\ 0.001),
  \quad k = K_g + 1, \ldots, K_{\max} \quad \text{(only when } K_g < K_{\max}\text{)}
\end{equation}

Without these priors, the unused parameters are unconstrained — they do not enter the likelihood, but the sampler still explores them. Pinning them to zero makes sampling much faster and has no effect on the reported parameters.

\subsection{S7.4 Sampler configuration}

\noindent\textbf{Table S7.2. Sampler settings for the primary model.}

\begingroup\small
\begin{longtable}{>{\raggedright\arraybackslash}p{0.30\linewidth}>{\raggedright\arraybackslash}p{0.60\linewidth}}
\toprule
\textbf{Setting} & \textbf{Value} \\
\midrule\endfirsthead
\toprule
\textbf{Setting} & \textbf{Value} \\
\midrule\endhead
Interface & cmdstanr 0.\allowbreak{}9.\allowbreak{}0 on CmdStan 2.\allowbreak{}38.\allowbreak{}0, R 4.\allowbreak{}6.\allowbreak{}1 \\
Chains & 4, run in parallel \\
Warmup iterations & 3,000 per chain \\
Sampling iterations & 4,000 per chain (16,000 post-warmup draws) \\
adapt\_\allowbreak{}delta & 0.\allowbreak{}99 \\
max\_\allowbreak{}treedepth & 14 \\
Seed & 42 \\
\bottomrule
\end{longtable}
\endgroup

\subsection{S7.5 Convergence}

We report diagnostics for the whole parameter vector, not just the parameters of interest.\allowbreak{}

\noindent\textbf{Table S7.3. Convergence diagnostics, primary model.}

\begingroup\small
\begin{longtable}{>{\raggedright\arraybackslash}p{0.20\linewidth}>{\raggedright\arraybackslash}p{0.18\linewidth}>{\raggedright\arraybackslash}p{0.52\linewidth}}
\toprule
\textbf{Diagnostic} & \textbf{Value} & \textbf{Reading} \\
\midrule\endfirsthead
\toprule
\textbf{Diagnostic} & \textbf{Value} & \textbf{Reading} \\
\midrule\endhead
Maximum R-hat, all parameters & 1.\allowbreak{}0024 & Below the 1.\allowbreak{}01 threshold \\
Minimum bulk ESS, all parameters & 640 & Attained by $\sigma_\gamma$ \\
Tail ESS, $\sigma_\gamma$ & 2,247 & Interval estimates are well resolved \\
Divergent transitions & 0 of 16,000 & No pathological geometry detected \\
Transitions at maximum tree depth & 4,004 of 16,000 (25\%) & Efficiency limit, not a validity failure \\
\bottomrule
\end{longtable}
\endgroup

Unlike divergent transitions, tree-depth saturation does not bias exploration; it only truncates trajectories early, costing efficiency.\allowbreak{} It follows directly from adapt\_\allowbreak{}delta = 0.\allowbreak{}99 on a 1,500-parameter posterior and is why the bulk ESS for $\sigma_\gamma$ is the lowest in the model (640) and the sampling run was long.\allowbreak{}

\subsection{S7.6 Posterior summary}

\noindent\textbf{Table S7.4. Posterior summary for the global parameters. Means are used elsewhere in the document; medians are added here for $\phi$, whose posterior is right-skewed.}

\begingroup\footnotesize
\begin{longtable}{>{\raggedright\arraybackslash}p{0.129\linewidth}>{\raggedright\arraybackslash}p{0.129\linewidth}>{\raggedright\arraybackslash}p{0.129\linewidth}>{\raggedright\arraybackslash}p{0.257\linewidth}>{\raggedright\arraybackslash}p{0.129\linewidth}>{\raggedright\arraybackslash}p{0.129\linewidth}}
\toprule
\textbf{Parameter} & \textbf{Mean} & \textbf{Median} & \textbf{90\% CI} & \textbf{R-hat} & \textbf{Bulk ESS} \\
\midrule\endfirsthead
\toprule
\textbf{Parameter} & \textbf{Mean} & \textbf{Median} & \textbf{90\% CI} & \textbf{R-hat} & \textbf{Bulk ESS} \\
\midrule\endhead
$\sigma_\gamma$ & 0.\allowbreak{}106 & 0.\allowbreak{}103 & [0.\allowbreak{}010, 0.\allowbreak{}219] & 1.\allowbreak{}002 & 640 \\
$\sigma_\beta$ & 0.\allowbreak{}249 & 0.\allowbreak{}249 & [0.\allowbreak{}210, 0.\allowbreak{}289] & 1.\allowbreak{}001 & 3,439 \\
$\phi$ & 1,102 & 966 & [564, 2,073] & 1.\allowbreak{}001 & 7,268 \\
\bottomrule
\end{longtable}
\endgroup

$\phi$ is large (posterior mean near 1,100), placing the likelihood close to multinomial: the field's composition is stable year to year.\allowbreak{} That signal-to-noise ratio is what makes a small $\sigma_\gamma$ detectable.\allowbreak{} $\sigma_\gamma$ is credibly above zero, with a 90\% interval whose lower bound sits close to it.\allowbreak{} $\sigma_\beta$ exceeds $\sigma_\gamma$ by a factor of 2.\allowbreak{}35, so post-2023 reshuffling is smaller than the variation the field has always shown.\allowbreak{}

None of the 241 analysed L3 methods has a $\gamma$ whose 90\% credible interval excludes zero.\allowbreak{} The aggregate signal does not localise to any named method; the directional lists in \S{}4 of the main text are ordered by posterior mean and are indicative only.\allowbreak{} Per-technique $\beta$ and $\gamma$ estimates are in \texttt{data/output/phi\_free/}.\allowbreak{}

\begin{center}
\begin{minipage}{\linewidth}
\centering
\includegraphics[width=\linewidth]{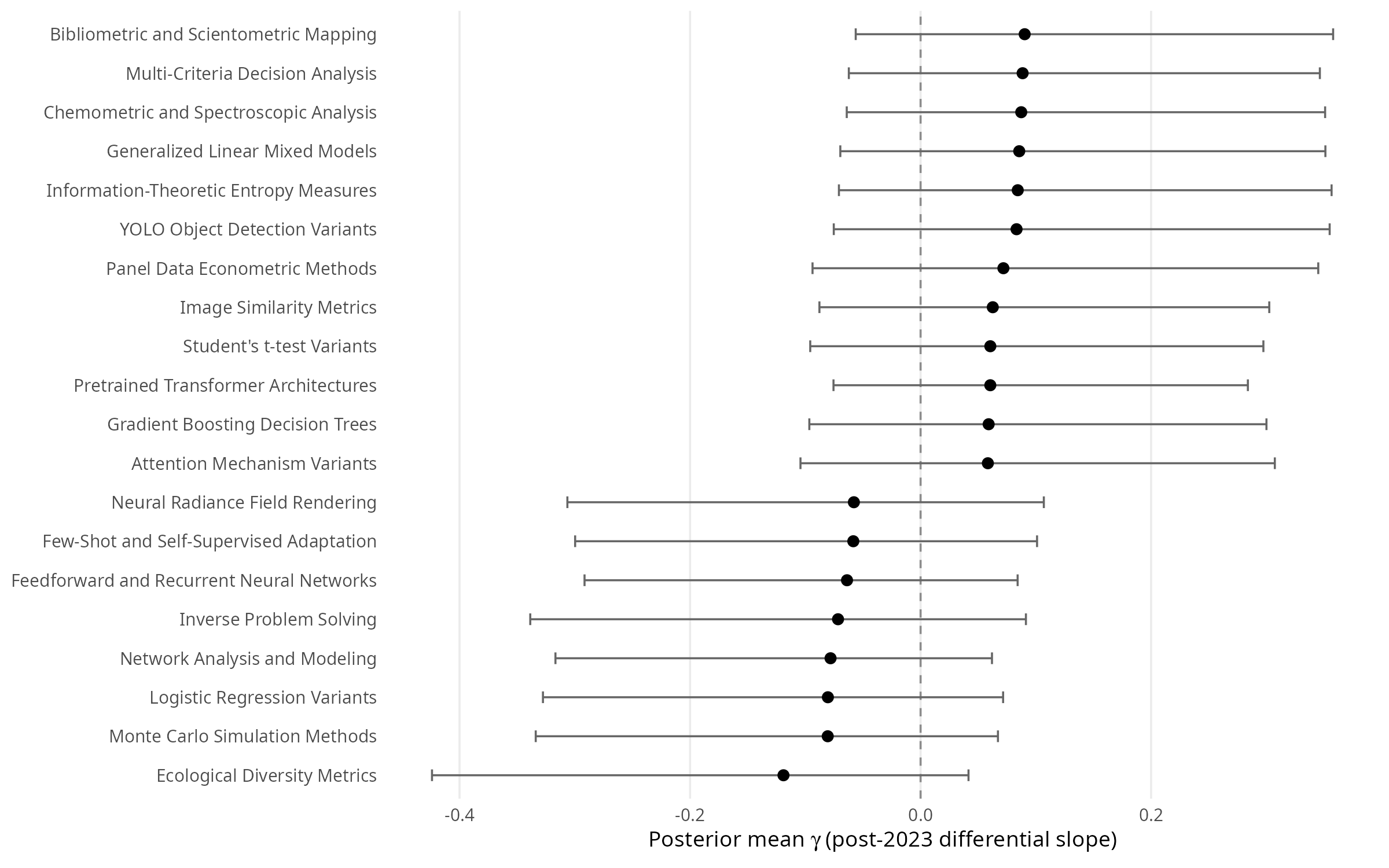}

\vspace{0.6em}
\noindent\textbf{Figure S7.1. Posterior $\gamma$ for the 20 L3 methods with the largest posterior mean $|\gamma|$.} $\gamma$ is the post-2023 differential slope, so these are the 20 methods furthest from zero. Bars are 90\% credible intervals. All 20 cross zero, as do the intervals of the other 221 methods.
\end{minipage}
\end{center}

\section{S8: Bayesian workflow checks}

We follow the Bayesian workflow of Gelman et al.\allowbreak{} (2020), running all checks from \texttt{R/bibliometric/04
\_workflow\_checks.R} against the fitted model; nothing is refitted except the fake-data simulation, which uses a single sub-discipline.\allowbreak{}

\subsection{S8.1 Prior predictive check}

We took 1,000 draws from the priors and pushed them through the model's generative structure for the five largest L2 sub-disciplines, without reference to the observed counts, with a randomly sampled year per draw. This checks whether the priors cover a plausible diversity range before the likelihood does any work.\allowbreak{}

The prior predictive 5th-95th percentile range for the inverse Simpson index is [2.\allowbreak{}76, 11.\allowbreak{}57], against an observed range of [1, 12.\allowbreak{}76].\allowbreak{} The priors cover the observed range without concentrating on it.\allowbreak{}

\subsection{S8.2 Posterior predictive check}

For each posterior draw and each group-year cell with data, we simulated replacement counts from the predicted proportions and recomputed the inverse Simpson index; 200 draws per cell.\allowbreak{} The tail probability per sub-discipline locates the observation within the predictive distribution: values near 0.5 indicate good calibration, values near 0 or 1 indicate systematic misfit.\allowbreak{}

All 25 of 25 sub-disciplines fall inside the 0.\allowbreak{}05-0.\allowbreak{}95 band.\allowbreak{} The model reproduces the compositional structure of every group it was fitted to.\allowbreak{}

\subsection{S8.3 Fake-data simulation}

Because $\sigma_\gamma$ is central to the analysis, we assess its identifiability directly. We simulated counts using known parameter values ($\sigma_\gamma = 0.060$ and $\sigma_\beta = 0.1$) and the data structure of the largest sub-discipline, and then refitted the model to the simulated data.

\noindent\textbf{Table S8.1. Fake-data recovery of $\sigma_\gamma$}

\begingroup\small
\begin{longtable}{>{\raggedright\arraybackslash}p{0.30\linewidth}>{\raggedright\arraybackslash}p{0.60\linewidth}}
\toprule
\textbf{Quantity} & \textbf{Value} \\
\midrule\endfirsthead
\toprule
\textbf{Quantity} & \textbf{Value} \\
\midrule\endhead
True $\sigma_\gamma$ & 0.\allowbreak{}060 \\
Recovered posterior mean & 0.\allowbreak{}168 \\
Recovered 90\% CI & [0.\allowbreak{}007, 0.\allowbreak{}451] \\
True value inside interval & Yes \\
\bottomrule
\end{longtable}
\endgroup

The recovered posterior is wide, which is expected: a single sub-discipline with three post-2023 years carries little information about a scale parameter.\allowbreak{} The posterior does contract toward the true value relative to the Exponential(4) prior, so $\sigma_\gamma$ is identified rather than prior-driven.\allowbreak{} In the full model the estimate is sharpened by aggregation across 25 sub-disciplines.\allowbreak{}

Cross-validation was not performed; we fitted only one model class, which is a limitation, though it weighs less here because the primary estimand is a scale parameter rather than a predictive quantity.\allowbreak{}

\section{S9: The concentration analysis}

Technical details for \S{}3.\allowbreak{}4.\allowbreak{}1; computed by \texttt{R/experiment/06\_concentration.R}.

\subsection{S9.1 Statistical method}

We model recommendation counts over the L3 taxonomy as multinomial.\allowbreak{} Under a uniform Dirichlet(1) prior the posterior is available in closed form:

\begin{equation}
\begin{aligned}
\mathbf{n} &\sim \operatorname{Multinomial}(\boldsymbol{\theta}), \\
\boldsymbol{\theta} &\sim \operatorname{Dirichlet}(\mathbf{1}), \\
\Rightarrow\ \boldsymbol{\theta}\mid\mathbf{n} &\sim \operatorname{Dirichlet}(\mathbf{1} + \mathbf{n}).
\end{aligned}
\end{equation}

No MCMC is needed; this is what we mean in the main text by ``computable exactly.'' With a single proportion vector per condition and no groups, years, or trend parameters, conjugacy applies. We draw samples via the standard gamma construction: $\Gamma(\alpha_k, 1)$ independently per category, then normalise.

The uniform prior adds one pseudo-count to every L3 category, including those a given condition never recommends. This is conservative: it inflates the apparent diversity of the most concentrated conditions, so the gap we report between LLM profiles and the literature is if anything understated.

For each of 4,000 draws, we compute the inverse Simpson index as $1/\sum_k \theta_k^2$, giving a full posterior distribution for the effective number of methods. Credible intervals are thus directly comparable across conditions.\allowbreak{}

\subsection{S9.2 Results}

\noindent\textbf{Table S9.1. Effective number of L3 methods by condition. Posterior means over 4,000 draws.}

\begingroup\small
\begin{longtable}{>{\raggedright\arraybackslash}p{0.20\linewidth}>{\raggedright\arraybackslash}p{0.18\linewidth}>{\raggedright\arraybackslash}p{0.52\linewidth}}
\toprule
\textbf{Condition} & \textbf{Effective methods} & \textbf{90\% CI} \\
\midrule\endfirsthead
\toprule
\textbf{Condition} & \textbf{Effective methods} & \textbf{90\% CI} \\
\midrule\endhead
Pre-2023 literature & 87.\allowbreak{}6 & [84.\allowbreak{}6, 90.\allowbreak{}6] \\
Post-2023 literature & 111.\allowbreak{}2 & [107.\allowbreak{}7, 114.\allowbreak{}7] \\
Qwen3 - overall & 31.\allowbreak{}6 & [30.\allowbreak{}2, 33.\allowbreak{}0] \\
Qwen3 - novice & 20.\allowbreak{}9 & [19.\allowbreak{}5, 22.\allowbreak{}4] \\
Qwen3 - intermediate & 30.\allowbreak{}6 & [28.\allowbreak{}3, 33.\allowbreak{}0] \\
Qwen3 - expert & 60.\allowbreak{}7 & [56.\allowbreak{}8, 64.\allowbreak{}7] \\
Gemma - overall & 28.\allowbreak{}8 & [27.\allowbreak{}6, 30.\allowbreak{}0] \\
Gemma - novice & 20.\allowbreak{}5 & [19.\allowbreak{}0, 21.\allowbreak{}9] \\
Gemma - intermediate & 29.\allowbreak{}5 & [27.\allowbreak{}3, 31.\allowbreak{}7] \\
Gemma - expert & 46.\allowbreak{}8 & [43.\allowbreak{}4, 50.\allowbreak{}2] \\
\bottomrule
\end{longtable}
\endgroup

Both models show the same two patterns.\allowbreak{} The literature diversified rather than contracted after 2023, and every LLM condition sits far below either literature figure, with no overlap of intervals.\allowbreak{} Within each model the ordering novice < intermediate < expert is strict; the expert profile reaches roughly half to two thirds of the pre-2023 literature's breadth without matching it.\allowbreak{} We treat the gradient's consistency across two models trained on different corpora as structural, though two models cannot establish that conclusively.\allowbreak{}

\begin{center}
\begin{minipage}{\linewidth}
\centering
\includegraphics[width=\linewidth]{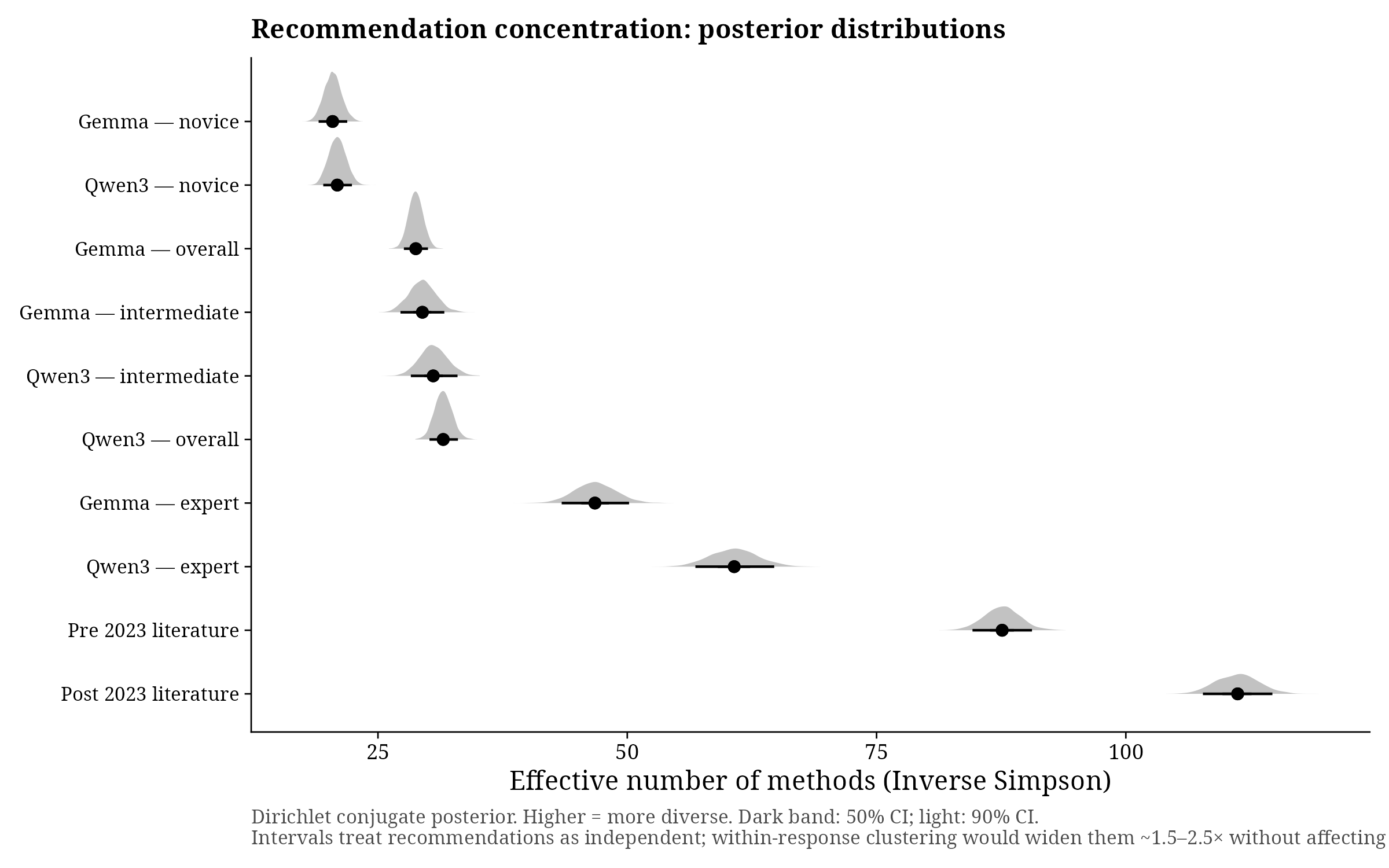}

\vspace{0.6em}
\noindent\textbf{Figure S9.1. Posterior distributions of the effective number of L3 methods (inverse Simpson index), by condition.} One density per row of Table~S9.1, each from 4{,}000 draws of the conjugate Dirichlet--multinomial posterior. Both literature posteriors sit to the right of every Qwen3 and Gemma profile, with no overlap; within each model, novice, intermediate and expert separate in that order.
\end{minipage}
\end{center}

\section{S10: What drives LLM recommendations}

Technical details for \S{}3.\allowbreak{}4.\allowbreak{}2; model at \texttt{stan/experiment\_prevalence\_nb.stan}, fitted by 

\texttt{R/experiment/07\_literature\_vs\_llm.R}.\allowbreak{}

\subsection{S10.1 Model}

\begin{equation}
n_{\mathrm{rec},i} \sim \operatorname{NegBinomial2}\!\left(\exp\left(\alpha + \beta_{\mathrm{pre}}\, x_{\mathrm{pre},i} + \beta_{\gamma}\, \gamma_{\mathrm{signed},i}\right),\ \phi\right)
\end{equation}

$n_{\mathrm{rec},i}$ is the number of times technique $i$ was recommended; $x_{\mathrm{pre},i}$ is log1p of its pre-2023 literature count, standing in for exposure in the training corpus; $\gamma_{\mathrm{signed},i}$ is its signed posterior mean $\gamma$ from the primary model.\allowbreak{} We use a negative binomial because the counts are strongly overdispersed: a few techniques are recommended very often and most rarely.\allowbreak{}

The two predictors are entered together.\allowbreak{} Pre-2023 prevalence and post-2023 movement are correlated: a regression on $\gamma$ alone would conflate a model reproducing its training distribution with a model tracking recent disciplinary change.\allowbreak{} Including both allows the mean-collapse mechanism to be tested rather than assumed.\allowbreak{}

\noindent\textbf{Table S10.1. Priors on the negative-binomial regression.}

\begingroup\small
\begin{longtable}{>{\raggedright\arraybackslash}p{0.30\linewidth}>{\raggedright\arraybackslash}p{0.60\linewidth}}
\toprule
\textbf{Parameter} & \textbf{Prior} \\
\midrule\endfirsthead
\toprule
\textbf{Parameter} & \textbf{Prior} \\
\midrule\endhead
$\alpha$ & $\operatorname{Normal}(0, 2)$ \\
$b_{\mathrm{pre}}$ & $\operatorname{Normal}(0, 1)$ \\
$b_\gamma$ & $\operatorname{Normal}(0, 1)$ \\
$\phi$ & $\operatorname{Exponential}(1)$ \\
\bottomrule
\end{longtable}
\endgroup

Fitted with 4 chains, 1,000 warmup and 1,000 sampling iterations each, adapt\_\allowbreak{}delta = 0.\allowbreak{}95, seed 42.\allowbreak{} Across all eight fits the maximum R-hat is 1.\allowbreak{}005 and the minimum effective sample size is 1,329.\allowbreak{}

\subsection{S10.2 Results}

\noindent\textbf{Table S10.2. Posterior means, 90\% credible intervals and posterior probability of exceeding zero, for the eight model-profile fits.}

\begingroup\footnotesize
\begin{longtable}{>{\raggedright\arraybackslash}p{0.1\linewidth}>{\raggedright\arraybackslash}p{0.1\linewidth}>{\raggedright\arraybackslash}p{0.1\linewidth}>{\raggedright\arraybackslash}p{0.1\linewidth}>{\raggedright\arraybackslash}p{0.1\linewidth}>{\raggedright\arraybackslash}p{0.1\linewidth}>{\raggedright\arraybackslash}p{0.2\linewidth}>{\raggedright\arraybackslash}p{0.1\linewidth}}
\toprule
\textbf{Model} & \textbf{Profile} & $\boldsymbol{b_{\mathrm{pre}}}$ & \textbf{90\% CI} & \textbf{P(>0)} & $\boldsymbol{b_\gamma}$ & \textbf{90\% CI} & \textbf{P(>0)} \\
\midrule\endfirsthead
\toprule
\textbf{Model} & \textbf{Profile} & $\boldsymbol{b_{\mathrm{pre}}}$ & \textbf{90\% CI} & \textbf{P(>0)} & $\boldsymbol{b_\gamma}$ & \textbf{90\% CI} & \textbf{P(>0)} \\
\midrule\endhead
Qwen3 & novice & 1.\allowbreak{}039 & [0.\allowbreak{}810, 1.\allowbreak{}275] & 1.\allowbreak{}00 & 0.\allowbreak{}089 & [-1.\allowbreak{}528, 1.\allowbreak{}656] & 0.\allowbreak{}53 \\
Qwen3 & intermediate & 0.\allowbreak{}675 & [0.\allowbreak{}484, 0.\allowbreak{}870] & 1.\allowbreak{}00 & 0.\allowbreak{}070 & [-1.\allowbreak{}543, 1.\allowbreak{}657] & 0.\allowbreak{}52 \\
Qwen3 & expert & 0.\allowbreak{}419 & [0.\allowbreak{}256, 0.\allowbreak{}578] & 1.\allowbreak{}00 & 0.\allowbreak{}212 & [-1.\allowbreak{}324, 1.\allowbreak{}743] & 0.\allowbreak{}60 \\
Qwen3 & overall & 0.\allowbreak{}657 & [0.\allowbreak{}509, 0.\allowbreak{}806] & 1.\allowbreak{}00 & 0.\allowbreak{}115 & [-1.\allowbreak{}391, 1.\allowbreak{}701] & 0.\allowbreak{}54 \\
Gemma & novice & 0.\allowbreak{}675 & [0.\allowbreak{}378, 0.\allowbreak{}984] & 1.\allowbreak{}00 & 0.\allowbreak{}131 & [-1.\allowbreak{}519, 1.\allowbreak{}738] & 0.\allowbreak{}56 \\
Gemma & intermediate & 0.\allowbreak{}543 & [0.\allowbreak{}332, 0.\allowbreak{}758] & 1.\allowbreak{}00 & 0.\allowbreak{}098 & [-1.\allowbreak{}504, 1.\allowbreak{}641] & 0.\allowbreak{}54 \\
Gemma & expert & 0.\allowbreak{}511 & [0.\allowbreak{}338, 0.\allowbreak{}685] & 1.\allowbreak{}00 & 0.\allowbreak{}235 & [-1.\allowbreak{}376, 1.\allowbreak{}849] & 0.\allowbreak{}59 \\
Gemma & overall & 0.\allowbreak{}576 & [0.\allowbreak{}400, 0.\allowbreak{}751] & 1.\allowbreak{}00 & 0.\allowbreak{}191 & [-1.\allowbreak{}354, 1.\allowbreak{}759] & 0.\allowbreak{}57 \\
\bottomrule
\end{longtable}
\endgroup

$b_{\mathrm{pre}}$ is credibly positive in all eight fits, with a posterior probability of exceeding zero of 1.\allowbreak{}00 throughout.\allowbreak{} The gradient runs in the predicted direction in both models: the coefficient is largest under the novice prompt and smallest under the expert prompt, so the less methodological guidance the prompt carries, the more closely recommendations track the pre-2023 canon.\allowbreak{} The gradient is considerably steeper for Qwen3 (1.\allowbreak{}039 to 0.\allowbreak{}419) than for Gemma (0.\allowbreak{}675 to 0.\allowbreak{}511).\allowbreak{}

$b_\gamma$ is indistinguishable from zero everywhere.\allowbreak{} Intervals span roughly $-1.5$ to $+1.8$ and the posterior probability of exceeding zero stays between 0.\allowbreak{}52 and 0.\allowbreak{}60, barely displaced from 0.\allowbreak{}50.\allowbreak{} This null result is not informative on its own and does not indicate absence of an effect.\allowbreak{} The predictor is built from the per-technique $\gamma$ estimates of the primary model, and none of those 241 estimates is individually credible (S7.\allowbreak{}6), so the regressor is dominated by measurement error.\allowbreak{} A null was the expected outcome regardless of the underlying truth.\allowbreak{} Sensitivity C addresses the same question by a route that does not depend on individual $\gamma$ values.\allowbreak{}

\section{S11: Robustness and sensitivity}

We performed five checks, each targeting a specific way the main findings could be an artefact of an analytical choice.\allowbreak{}

\subsection{S11.1 Breakpoint at 2022}

The 2023 breakpoint is a judgement about when ChatGPT entered serious academic use.\allowbreak{} Refitting with post\_\allowbreak{}llm set from 2022, ChatGPT's launch year, tests whether the result depends on it.\allowbreak{}

\noindent\textbf{Table S11.1. Primary model under both breakpoints.}

\begingroup\small
\begin{longtable}{>{\raggedright\arraybackslash}p{0.20\linewidth}>{\raggedright\arraybackslash}p{0.18\linewidth}>{\raggedright\arraybackslash}p{0.52\linewidth}}
\toprule
\textbf{Quantity} & \textbf{2023 breakpoint} & \textbf{2022 breakpoint} \\
\midrule\endfirsthead
\toprule
\textbf{Quantity} & \textbf{2023 breakpoint} & \textbf{2022 breakpoint} \\
\midrule\endhead
$\sigma_\gamma$ & 0.\allowbreak{}106 [0.\allowbreak{}010, 0.\allowbreak{}219] & 0.\allowbreak{}044 [0.\allowbreak{}004, 0.\allowbreak{}109] \\
$\sigma_\beta$ & 0.\allowbreak{}249 [0.\allowbreak{}210, 0.\allowbreak{}289] & 0.\allowbreak{}254 [0.\allowbreak{}218, 0.\allowbreak{}294] \\
Techniques with credible $\gamma$ & 0 of 241 & 0 of 241 \\
Correlation of per-technique $\gamma$ & -- & 0.\allowbreak{}769 \\
\bottomrule
\end{longtable}
\endgroup

The qualitative conclusions are unchanged: $\sigma_\beta$ is stable, no individual technique reaches credibility under either specification, and per-technique $\gamma$ estimates correlate at 0.\allowbreak{}77.\allowbreak{} $\sigma_\gamma$ is smaller under the 2022 break, which is the expected direction.\allowbreak{} Moving the break a year earlier assigns 2022, a year with no plausible LLM effect, to the post-treatment period, diluting the contrast.\allowbreak{} The result is consistent with a shift located at 2023 rather than 2022.\allowbreak{}

\subsection{S11.2 Sensitivity A: count threshold}

If noisy $\gamma$ estimates for rarely used techniques attenuate the coefficient toward zero, restricting to well-represented techniques should sharpen it. We restrict to methods with at least 50 papers in 2023--2025, leaving 33 of 242.\allowbreak{} The denominator here and in S11.\allowbreak{}3 is the full 242-label taxonomy rather than the 241 analysed clusters, because both checks operate on the pre-filter vocabulary; see S12.\allowbreak{}1.\allowbreak{} The overall coefficient is $-0.138$ with $P(\beta > 0) = 0.438$: still null, and still slightly negative.\allowbreak{} The null is not an artefact of rare methods.\allowbreak{}

\subsection{S11.3 Sensitivity B: taxonomy remapping}

An earlier experiment run was classified against taxonomy v2 (225 L3 methods), and only 53 of 242 v3 methods matched by direct label comparison.\allowbreak{} Remapping v2 onto v3 by prefix, exact and fuzzy string matching raises coverage to 205 of 242 (85\%).\allowbreak{} The result is unchanged: overall $\beta = -0.339$, $P(\beta > 0) = 0.368$, with all profiles negative.\allowbreak{} The current experiment data were classified directly against v3, so this check applies only to the earlier run.\allowbreak{}

\subsection{S11.4 Sensitivity C: distributional test}

This is the most informative of the five, because it asks the Step 3 question without depending on individual $\gamma$ estimates.\allowbreak{} Rather than regressing recommendation counts on noisy per-technique $\gamma$ values, we compare whole frequency distributions: if LLMs are implicated in post-2023 change, their recommendation distribution should resemble the post-2023 literature more than the pre-2023 literature.\allowbreak{}

\noindent\textbf{Table S11.2. Distributional similarity of LLM recommendations to the pre- and post-2023 literature.}

\begingroup\small
\begin{longtable}{>{\raggedright\arraybackslash}p{0.30\linewidth}>{\raggedright\arraybackslash}p{0.60\linewidth}}
\toprule
\textbf{Quantity} & \textbf{Value} \\
\midrule\endfirsthead
\toprule
\textbf{Quantity} & \textbf{Value} \\
\midrule\endhead
Delta cosine similarity (post minus pre) & +0.\allowbreak{}071 \\
90\% CI & [+0.\allowbreak{}058, +0.\allowbreak{}085] \\
P(delta > 0) & 1.\allowbreak{}000 \\
Permutation test & p = 0.\allowbreak{}0013 \\
Novice /\allowbreak{} intermediate /\allowbreak{} expert delta & +0.\allowbreak{}080 /\allowbreak{} +0.\allowbreak{}069 /\allowbreak{} +0.\allowbreak{}013 \\
\bottomrule
\end{longtable}
\endgroup

The recommendation distribution is credibly closer to the post-2023 literature, and the profile gradient matches the mean-collapse prediction, with the weakest guidance producing the strongest alignment.\allowbreak{} This recovers a signal the regression could not detect, for the reason given in S10.\allowbreak{}2.\allowbreak{}

The test establishes distributional similarity, not causation.\allowbreak{} Recommendations could resemble the post-2023 literature because models influenced adoption, because they mirror trends already present in their training data, or both, and this design cannot separate those.\allowbreak{}

\subsection{S11.5 Sensitivity D: direct diversity trajectory}

The primary model works at the technique level. This check asks the simpler question directly: did diversity within sub-disciplines change after 2023? We take posterior draws of the inverse Simpson index from the primary model's generated quantities block as input to a second-level model with the same two-slope structure, propagating the upstream posterior standard deviation as measurement error.\allowbreak{}

\noindent\textbf{Table S11.3. Diversity-trajectory model (Sensitivity D).}

\begingroup\small
\begin{longtable}{>{\raggedright\arraybackslash}p{0.20\linewidth}>{\raggedright\arraybackslash}p{0.18\linewidth}>{\raggedright\arraybackslash}p{0.52\linewidth}}
\toprule
\textbf{Parameter} & \textbf{Estimate} & \textbf{90\% CI} \\
\midrule\endfirsthead
\toprule
\textbf{Parameter} & \textbf{Estimate} & \textbf{90\% CI} \\
\midrule\endhead
$\sigma_\gamma$ & 0.\allowbreak{}064 & [0.\allowbreak{}003, 0.\allowbreak{}174] \\
$\sigma_\beta$ & 0.\allowbreak{}677 & [0.\allowbreak{}510, 0.\allowbreak{}910] \\
\bottomrule
\end{longtable}
\endgroup

All 25 sub-discipline $\gamma$ intervals straddle zero and $\sigma_\beta$ exceeds $\sigma_\gamma$ by an order of magnitude.\allowbreak{} Read together with the primary model, the field reorients internally without measurably homogenising at the sub-discipline level.\allowbreak{}

\section{S12: Reproducibility}

\subsection{S12.1 Corpus flow into the model}

S3.\allowbreak{}3 traces the corpus to the 8,398 articles that survive removal of the garbage cluster.\allowbreak{} The modelling stream applies one further filter, and the table below carries the chain to the array passed to Stan.\allowbreak{} Every figure was re-derived from the deposited artefacts by the verification script.\allowbreak{}

The early stages of the pipeline depend on Scopus exports and are not reproducible without a Scopus subscription. The deposited \texttt{stan\_data.rds} (built by \texttt{00\_data\_prep.R}) contains the aggregated count array with no article-level metadata; every analysis from \S{}3.\allowbreak{}3 onward can be reproduced from that file alone.

\noindent\textbf{Table S12.1. Corpus flow into the Dirichlet-Multinomial model. Sources marked $\dagger$ are Scopus-derived and not deposited; sources marked $\ddagger$ are deposited.}

\begingroup\small
\begin{longtable}{>{\raggedright\arraybackslash}p{0.20\linewidth}>{\raggedright\arraybackslash}p{0.18\linewidth}>{\raggedright\arraybackslash}p{0.52\linewidth}}
\toprule
\textbf{Stage} & \textbf{Count} & \textbf{Source} \\
\midrule\endfirsthead
\toprule
\textbf{Stage} & \textbf{Count} & \textbf{Source} \\
\midrule\endhead
Cleaned corpus & 119,327 & \texttt{df\_cleaned.xlsx}$^\dagger$ \\
Articles with at least one method & 8,404 & \texttt{taxonomy\_abstract\_join.csv}$^\dagger$ \\
(article, method) pairs & 17,229 & \texttt{taxonomy\_abstract\_join.csv}$^\dagger$ \\
Articles after the garbage cluster & 8,398 & 145 pairs, 6 articles removed \\
Articles entering the model & 7,355 & after the 2010--2025 year filter \\
Paper-method observations & 14,138 & \texttt{stan\_data.rds}$^\ddagger$, sum of count array \\
L3 techniques analysed & 241 of 242 & one cluster discarded at Phase 4.\allowbreak{}5 \\
L2 sub-disciplines & 25 & no singleton groups, none dropped \\
\bottomrule
\end{longtable}
\endgroup

The year filter removes 1,043 articles: 403 dated 2026 and 640 dated before 2010.\allowbreak{} 2026 is excluded as a partial year, since retrieval took place in April 2026 and Scopus indexing lags publication; including it would depress the final year of every trajectory for reasons that have nothing to do with method choice.\allowbreak{} The 14,138 observations decompose as 8,442 pre-2023 and 5,696 post-2023.\allowbreak{}

Two counts are easily confused and are kept distinct throughout.\allowbreak{} 242 is the number of L3 clusters the taxonomy produced; 241 is the number carried into analysis.\allowbreak{} Where the main text reports coverage of the taxonomy by LLM recommendations, the denominator is 242, because both models mapped some items into the discarded cluster.\allowbreak{}

\subsection{S12.2 Software environment}

\noindent\textbf{Table S12.2. Software for the modelling stream. The Python environment for the machine-learning stream is given in S1.}

\begingroup\small
\begin{longtable}{>{\raggedright\arraybackslash}p{0.30\linewidth}>{\raggedright\arraybackslash}p{0.60\linewidth}}
\toprule
\textbf{Component} & \textbf{Version} \\
\midrule\endfirsthead
\toprule
\textbf{Component} & \textbf{Version} \\
\midrule\endhead
R & 4.\allowbreak{}6.\allowbreak{}1 \\
cmdstanr & 0.\allowbreak{}9.\allowbreak{}0 \\
CmdStan & 2.\allowbreak{}38.\allowbreak{}0 \\
Random seed, all fits & 42 \\
\bottomrule
\end{longtable}
\endgroup

\subsection{S12.3 Scripts and deposited artefacts}

Scripts are run with Rscript from the project root, in this order:

R/\allowbreak{}bibliometric/\allowbreak{}00\_\allowbreak{}data\_\allowbreak{}prep.\allowbreak{}R - builds stan\_\allowbreak{}data.\allowbreak{}rds and vocab.\allowbreak{}rds

R/\allowbreak{}bibliometric/\allowbreak{}01\_\allowbreak{}fit\_\allowbreak{}dm\_\allowbreak{}model.\allowbreak{}R - fits the primary model

R/\allowbreak{}bibliometric/\allowbreak{}02\_\allowbreak{}extract\_\allowbreak{}plot.\allowbreak{}R - posterior extraction and figures

R/\allowbreak{}bibliometric/\allowbreak{}03\_\allowbreak{}extract\_\allowbreak{}gamma.\allowbreak{}R - per-technique gamma table

R/\allowbreak{}bibliometric/\allowbreak{}04\_\allowbreak{}workflow\_\allowbreak{}checks.\allowbreak{}R - the checks in S8

R/\allowbreak{}bibliometric/\allowbreak{}05\_\allowbreak{}robustness\_\allowbreak{}2022.\allowbreak{}R - the 2022 breakpoint refit

R/\allowbreak{}experiment/\allowbreak{}06\_\allowbreak{}concentration.\allowbreak{}R - the concentration analysis

R/\allowbreak{}experiment/\allowbreak{}07\_\allowbreak{}literature\_\allowbreak{}vs\_\allowbreak{}llm.\allowbreak{}R - the negative-binomial regression

R/\allowbreak{}sensitivity/\allowbreak{} - Sensitivities C and D

The raw Scopus metadata and the 546 source identifiers are proprietary to Elsevier and are not redistributed; S3.\allowbreak{}4 gives the procedure for regenerating the identifier list.\allowbreak{} Scripts \texttt{00} and \texttt{01} in the bibliometric stream therefore require Scopus access to run from scratch. Everything they produce is deposited: \texttt{stan\_data.rds} (the aggregated count array), the full taxonomy with member terms and frequencies, the complete stimulus set, every recommendation produced by both models with its justification and all three mapping votes, the Stan models, the fitted posterior summaries, and the analysis scripts.\allowbreak{} A reader without Scopus access can start from \texttt{stan\_data.rds} and reproduce every analysis reported in \S{}4.\allowbreak{}

\section{List of supplementary tables}

For practical cross-reference to the information in the manuscript, we provide here a list of the tables contained in this ESM.

\begingroup\small
\begin{longtable}{>{\raggedright\arraybackslash}p{0.12\linewidth}>{\raggedright\arraybackslash}p{0.10\linewidth}>{\raggedright\arraybackslash}p{0.68\linewidth}}
\toprule
\textbf{Table} & \textbf{Section} & \textbf{Contents} \\
\midrule\endfirsthead
\toprule
\textbf{Table} & \textbf{Section} & \textbf{Contents} \\
\midrule\endhead
S1.1 & S1 & Computational environment for the machine-learning stream \\
S2.1 & S2 & Models and decoding settings, by pipeline stage \\
S3.2 & S3.2 & Scopus API metadata fields and their columns in the export \\
S3.3 & S3.3 & From raw Scopus records to the analysis corpus \\
S4.4 & S4.4 & Semantic embedding and L3 clustering settings \\
S5.1 & S5.1 & The 25 L2 categories, with definitions and corpus counts \\
S5.2 & S5.2 & The 241 L3 clusters, ordered by mentions \\
S6.1 & S6.1 & The 28 research questions, one per thematic category \\
S6.2 & S6.2 & The nine vague methodological descriptors \\
S6.6a & S6.6 & \texorpdfstring{L4$\rightarrow$L3}{L4->L3} mapping consistency, by model \\
S6.6b & S6.6 & L4 method items by researcher profile \\
S7.1 & S7.2 & Priors on the Dirichlet-Multinomial model \\
S7.2 & S7.4 & Sampler settings for the primary model \\
S7.3 & S7.5 & Convergence diagnostics, primary model \\
S7.4 & S7.6 & Posterior summary for the global parameters \\
S8.1 & S8.3 & Fake-data recovery of $\sigma_\gamma$ \\
S9.1 & S9.2 & Effective number of L3 methods by condition \\
S10.1 & S10.1 & Priors on the negative-binomial regression \\
S10.2 & S10.2 & Posterior means, 90\% credible intervals and $P(>0)$ for the eight model-profile fits \\
S11.1 & S11.1 & Primary model under both breakpoints \\
S11.2 & S11.4 & Distributional similarity of LLM recommendations to the pre- and post-2023 literature \\
S11.3 & S11.5 & Diversity-trajectory model (Sensitivity D) \\
S12.1 & S12.1 & Corpus flow into the Dirichlet-Multinomial model \\
S12.2 & S12.2 & Software for the modelling stream \\
\bottomrule
\end{longtable}
\endgroup

\section{List of supplementary figures}

\begingroup\small
\begin{longtable}{>{\raggedright\arraybackslash}p{0.12\linewidth}>{\raggedright\arraybackslash}p{0.10\linewidth}>{\raggedright\arraybackslash}p{0.68\linewidth}}
\toprule
\textbf{Figure} & \textbf{Section} & \textbf{Contents} \\
\midrule\endfirsthead
\toprule
\textbf{Figure} & \textbf{Section} & \textbf{Contents} \\
\midrule\endhead
S7.1 & S7.6 & Posterior $\gamma$ for the 20 L3 methods with the largest $|\gamma|$, with 90\% credible intervals \\
S9.1 & S9.2 & Posterior distributions of the effective number of L3 methods, by condition \\
\bottomrule
\end{longtable}
\endgroup

\end{document}